\documentstyle[makeidx,fleqn,11pt,epsfig]{article}
\input{amssym.def}
\input{amssym.tex}
\author{ }
\date{ }
\makeatletter

\def\gsim{\compoundrel>\over\sim}
\def\lsim{\compoundrel<\over\sim}
\def\compoundrel#1\over#2{\mathpalette\compoundreL{{#1}\over{#2}}}
\def\compoundreL#1#2{\compoundREL#1#2}
\def\compoundREL#1#2\over#3{\mathrel
      {\vcenter{\hbox{$\m@th\buildrel{#1#2}\over{#1#3}$}}}}
\begin{document}
%
\vspace*{0.5cm}
\hfill\ {\Large\bf KEK Preprint 2012-21}

\vspace*{2mm}
\hfill\ {\Large\bf August 2012~~~~~~~~~~~~~\,}

\vspace*{2mm}
\hfill\ {\Large\bf A/H\,~~~~~~~~~~~~~~~~~~~~~~~\,}

\vspace*{2.0cm}
\begin{center}
{\Huge A~~CLIC-Prototype~~Higgs~~Factory}\\ 
\vspace*{4.4cm}
{\Large R. BELUSEVIC~~and~~T. HIGO }\\
\vspace*{0.8cm}
{\large {\em High Energy Accelerator Research Organization} (KEK)}\\
\vspace*{1.3mm}
{\large 1-1 {\em Oho, Tsukuba, Ibaraki} 305-0801, {\em Japan}} \\
\vspace*{1.3mm}
{\large belusev@post.kek.jp}\\
\end{center}

\thispagestyle{empty}

\newpage

\tableofcontents
\addtocontents{toc}{\protect\vspace{1.3cm}}
\vspace*{5mm}
\noindent
{\large\bf References}


\newpage

\vspace*{0.5cm}
\begin{center}
\begin{minipage}[t]{13.6cm}
{\bf Abstract\,:}\hspace*{3mm}
{We propose that a pair of electron linacs with high accelerating gradients
and an optical FEL be built at an existing laboratory. 
The linacs would employ CLIC-type rf cavities and a klystron-based power 
source; a two-beam scheme could be implemented at a later stage. The proposed
facility would serve primarily as an $e^{+}e^{-}/\gamma\gamma$ Higgs-boson
factory. The rich set of final states in $e^{+}e^{-}$ and $\gamma\gamma$
collisions would play an essential role in measuring the mass, spin, parity, 
two-photon width and trilinear self-coupling of the Higgs boson, as well as its
couplings to fermions and gauge bosons. These quantities are more difficult to 
determine with only one initial state. For some processes within and beyond the
Standard Model, the required CM energy is considerably lower at the proposed
facility than at an $e^{+}e^{-}$ or proton collider.} 
\end{minipage}
\end{center}

\vspace*{0.5cm}
\renewcommand{\thesection}{\arabic{section}}
\section{~The Standard Model of particle physics}
\vspace*{0.3cm}

\setcounter{equation}{0}

~~~~Enormous progress has been made in the field of high-energy physics over 
the past five decades. The existence of a subnuclear world of quarks and 
leptons, whose dynamics can be described by quantum field theories possessing
gauge symmetry ({\em gauge theories}), has been firmly established. The {\em
Standard Model} (SM) of particle physics gives a coherent quantum-mechanical 
description of electromagnetic, weak and strong interactions based on
fundamental constituents --- quarks and leptons --- interacting via force 
carriers --- photons, W and Z bosons, and gluons. 

The Standard Model is supported by two theoretical `pillars': the {\em gauge
principle} and the {\em Higgs mechanism} for particle mass generation. Whereas 
the former has been firmly established through precision electroweak 
measurements, the latter is yet to be fully tested.

In the SM, where electroweak symmetry is broken by the Higgs mechanism, the
mass of a particle depends on its interaction with the Higgs field, a medium
that permeates the universe. The photon and the gluon do not have such
couplings, and so they remain massless. The Standard Model predicts the 
existence of a neutral spin-0 particle associated with the Higgs field, but it 
does not predict its mass. Although the existence of a Higgs field provides a
simple mechanism for electroweak symmetry breaking, {\em our inability to
predict the mass of the Higgs boson reflects the fact that we really do not 
understand at a fundamental level why this phenomenon occurs}. Another 
undesirable feature of the Standard Model is the {\em ad hoc} way in which 
fermion masses are introduced. 

\begin{figure}[t]
\vspace{-3.3mm}
\begin{center}
\epsfig{file=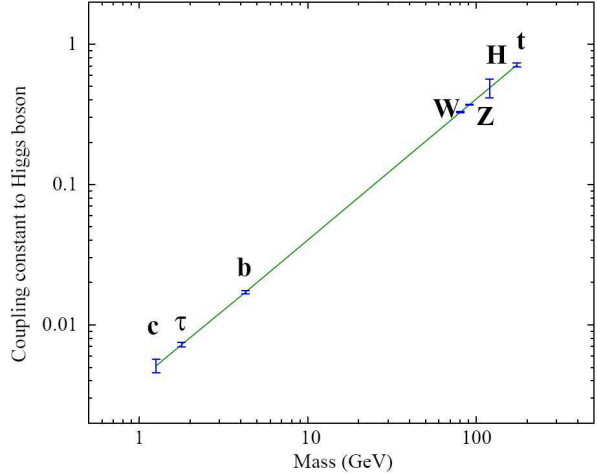,height=0.26\textheight}
\end{center}
\vskip -8mm
\caption{Precision with which the couplings of the Higgs particle with
$\mbox{\large{$m$}}_{\mbox{\tiny{H}}}^{~} = 120$ GeV can be determined at an
$e^{+}e^{-}$ collider with $\int\!L = 500~{\rm fb}^{-1}$. The coupling
$\kappa_{i}$ of the particle $i$ with mass $m_{i}$ is defined so that the
relation $m_{i} = v\kappa_{i}$ with $v \simeq 246$ GeV holds in the SM
\cite{GLC}.}
\label{fig:Higgs_couplings}
\end{figure}

All of the couplings of the Higgs particle to gauge bosons and fermions are
completely determined in the Standard Model in terms of electroweak coupling 
constants and fermion masses. Higgs production and decay processes can be 
computed in the SM unambiguously in terms of the Higgs mass alone. Since the
coupling of the Higgs boson to fermions and gauge bosons is proportional to the
particle masses (see Fig.\,\ref{fig:Higgs_couplings}), the Higgs boson will be
produced in association with heavy particles and will decay into the heaviest
particles that are kinematically accessible.
 
The rich set of final states in $e^{+}e^{-}$ and $\gamma\gamma$ collisions 
would play an essential role in measuring the mass ($\mbox{\large{$m$}}_{\mbox
{\tiny{H}}}^{~}$), spin, parity, two-photon width and trilinear self-coupling
of the Higgs boson, as well as its couplings to fermions and gauge bosons;
these quantities are more difficult to determine with only one initial state.

The Higgs-boson mass affects the values of electroweak observables 
through radiative corrections. The precision electroweak data obtained over the
past three decades consists of over a thousand individual measurements. Many of 
those measurements may be combined to provide a global test of consistency with
the SM. The best constraint on $\mbox{\large{$m$}}_{\mbox{\tiny{H}}}^{~}$ is
obtained by making a global fit to the electroweak data. Such a fit strongly 
suggests that the most likely mass for the SM Higgs boson is just above the 
limit of 114.4 GeV set by direct searches at the LEP $e^{+}e^{-}$ collider
\cite{LEP}.

\begin{figure}[t]
\begin{center}
\epsfig{file=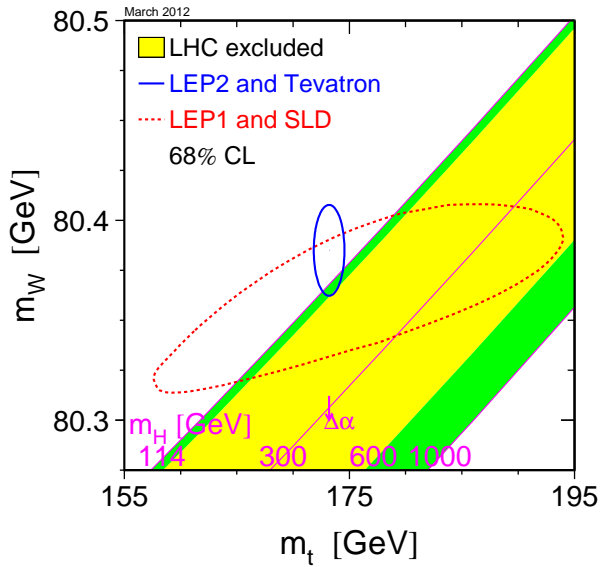,height=0.36\textheight}
\end{center}
\vskip -8mm
\caption{Direct and indirect constraints on the W and top-quark masses.
Reprinted courtesy of LEP Electroweak Working Group (Winter 2012).}
\label{fig:Mt-Mw}
\end{figure}

The solid ellipse in Fig.\,\ref{fig:Mt-Mw} indicates the direct measurement of
the W mass, $\mbox{\large{$m$}}_{\mbox{\tiny{W}}}^{~}$, and the top-quark mass,
$\mbox{\large{$m$}}_{t}^{~}$. The dashed ellipse represents the indirect 
constraints between the two masses. Also shown is the correlation between
$\mbox{\large{$m$}}_{\mbox{\tiny{W}}}^{~}$ and $\mbox{\large{$m$}}_{t}^{~}$ as
expected in the Standard Model for different values of the Higgs-boson mass
$\mbox{\large{$m$}}_{\mbox{\tiny{H}}}^{~}$ (the diagonal band). Notice that the
two ellipses overlap near the lines of constant $\mbox{\large{$m$}}_{\mbox
{\tiny{H}}}^{~}$. This indicates that the Standard Model is a fairly good
approximation to reality. Both ellipses are consistent with a low value of the
Higgs-boson mass.

High-precision electroweak measurements, therefore, provide a natural 
complement to direct studies of the Higgs sector. All the measurements made at
LEP and SLC could be repeated at the proposed facility using 90\% polarized
electron beams and at much higher luminosities. Assuming a geometric 
luminosity $L_{e^{+}e^{-}}^{~} \approx 5\times 10^{33}$ cm$^{-2}$\,s$^{-1}$ at
the Z resonance, about $2\times 10^{9}$ Z bosons can be produced in an 
operational year of $10^{7}$ s. This is about 200 times the entire LEP 
statistics. Moreover, about $10^{6}$ W bosons can be produced near the W-pair
threshold at the optimal energy point for measuring the W-boson mass. An 
increase in the number of Z events by two orders of magnitude as compared to 
LEP data, and a substantially improved accuracy in the measurement of W-boson
properties, would provide new opportunities for high-precision electroweak 
studies \cite{erler}.

\vspace*{0.3cm}
\renewcommand{\thesection}{\arabic{section}}
\section{~The Higgs mechanism}
\vspace*{0.3cm}

~~~~In order to provide a mechanism for the generation of particle masses in
the {\em Standard Model} without violating its gauge invariance, a
complex scalar SU(2) doublet $\Phi$ with four real fields and {\em hypercharge}
${\rm Y} = 1$ is introduced. The dynamics of the field $\Phi$ is described by
the Lagrangian
\begin{equation}
{\cal L}_{\Phi} \,=\, (D_{\mu}\Phi )^{\dag}(D^{\mu}\Phi ) \,-\, \mu^{2\,}
\Phi^{\dag}\Phi \,-\, \lambda\mbox{\large{$($}}\Phi^{\dag}\Phi\mbox{\large
{$)$}}^{2}
\end{equation}
where $(D_{\mu}\Phi )^{\dag}(D^{\mu}\Phi )$ is the kinetic-energy term and 
$\mu^{2\,}\Phi^{\dag}\Phi + \lambda\mbox{\large{$($}}\Phi^{\dag}\Phi\mbox
{\large{$)$}}^{2}$ is the Higgs self-interaction potential. In the so-called
{\em unitary gauge},
\begin{equation}
\Phi \,=\, \frac{\raisebox{-.3ex}{\mbox{\small{1}}}}{\sqrt{\mbox{\small
{2}}}}\left (\!\!
\begin{array}{c}
0 \\*[0.7mm]
v + {\rm H}
\end{array} \!\right )
\end{equation}
where $v \equiv \sqrt{-\mu^{2}/\lambda} = 246~{\rm GeV}$ is the {\em vacuum
expectation value} of the scalar field $\Phi$. The Higgs self-interaction
potential gives rise to terms involving only the physical {\em Higgs field} H: 
\begin{equation}
V_{\mbox{\tiny{H}}} \,=\, \frac{\raisebox{-.3ex}{\mbox{\small{1}}}}
{\raisebox{.2ex}{\mbox{\small{2}}}}\mbox{\Large{$($}}\mbox{\small{2}}\lambda
v^{2}\mbox{\Large{$)$}}{\rm H}^{2} \,+\, \lambda v\,{\rm H}^{3} \,+\, \frac
{\raisebox{-.3ex}{$\lambda$}}{\raisebox{.2ex}{\mbox{\small{4}}}}\,{\rm H}^{4}
\end{equation}

We see from Eq. (3) that the {\em Higgs mass} $\mbox{\large{$m$}}_{\mbox
{\tiny{H}}}^{~} \,=\, \sqrt{\mbox{\small{2}}\lambda}\,v$ is related to the
quadrilinear self-coupling strength $\lambda$. It is also evident that the
{\em trilinear self-coupling} of the Higgs field is 
\begin{equation}
\lambda_{\mbox{\tiny{HHH}}} \,\equiv\, \lambda v \,=\, \frac{\mbox{\large{$m$}}
_{\mbox{\tiny{H}}}^{\,2}}{\raisebox{.3ex}{\mbox{\small{2}}$v$}}
\label{eq:hhh}
\end{equation}
and the self-coupling among four Higgs fields
\begin{equation}
\lambda_{\mbox{\tiny{HHHH}}} \,\equiv\, \frac{\raisebox{-.3ex}{$\lambda$}}
{\raisebox{.2ex}{\mbox{\small{4}}}} \,=\, \frac{\mbox{\large{$m$}}_{\mbox{\tiny
{H}}}^{\,2}}{\mbox{\small{8}}v^{2}}
\end{equation}
Note that the Higgs self-couplings are uniquely determined by the mass of the 
Higgs boson, which represents a free parameter of the model.

Any theoretical model based on the gauge principle must evoke spontaneous
symmetry breaking. For example, the {\em minimal supersymmetric} extension of
the Standard Model ({\small MSSM}) introduces two SU(2) doublets of complex
Higgs fields, whose neutral components have vacuum expectation values 
$v_{\mbox{\tiny{1}}}$ and $v_{\mbox{\tiny{2}}}$. In this model,
spontaneous electroweak symmetry breaking results in five physical
Higgs-boson states: two neutral scalar fields $h^0$ and $H^0$, a
pseudoscalar $A^0$ and two charged bosons $H^\pm$. This extended Higgs
system can be described at tree level by two parameters: the ratio
$\tan\beta \equiv v_{\mbox{\tiny{2}}}/v_{\mbox{\tiny{1}}}$, and a mass
parameter, which is generally identified with the mass of the
pseudoscalar boson $A^0$, $\mbox{\large{$m$}}_{\mbox{\tiny{A}}}^{~}$. While 
there is a bound of about 140~GeV on the mass of the lightest CP-even neutral
Higgs boson $h^0$ \cite{h0-mass1,h0-mass2}, the masses of the $H^0$, $A^0$ and
$H^\pm$ bosons may be much larger. The existence of the Higgs boson $h^0$
is the only verifiable low-energy prediction of the MSSM model.

The trilinear self-coupling of the lightest MSSM Higgs boson at tree level is
given by
\begin{equation}
\lambda_{hhh}  \,=\, \frac{\mbox{\large{$m$}}_{\mbox{\tiny{Z}}}^{\,2}}
{\raisebox{.3ex}{\mbox{\small{2}}$v$}}\cos 2\alpha\sin(\beta+\alpha)
\hspace*{10mm}{\rm where}\hspace*{10mm}\tan 2\alpha \,=\, \tan 2\beta \,\frac
{\mbox{\large{$m$}}_{\mbox{\tiny{$A$}}}^{\,2} + \mbox{\large{$m$}}_{\mbox{\tiny
{Z}}}^{\,2}}{\mbox{\large{$m$}}_{\mbox{\tiny{$A$}}}^{\,2} - \mbox{\large{$m$}}_
{\mbox{\tiny{Z}}}^{\,2}}
\end{equation}
We see that for arbitrary values of the MSSM input parameters $\tan\beta$ and
$\mbox{\large{$m$}}_{\mbox{\tiny{$A$}}}^{~}$ the value of the $h^0$
self-coupling differs from that of the SM Higgs boson. However, in the
so-called `decoupling limit' $\mbox{\large{$m$}}_{\mbox{\tiny{$A$}}}^{\,2} \sim
\mbox{\large{$m$}}_{\mbox{\tiny{$H$}}^{0}}^{\,2} \sim \mbox{\large{$m$}}_{\mbox
{\tiny{$H$}}^{\pm}}^{\,2} \gg v^2/2$, the trilinear and quadrilinear 
self-couplings of the lightest CP-even neutral Higgs boson $h^0$ approach the
SM value. 

In contrast to any anomalous couplings of the gauge bosons, an anomalous 
self-coupling of the Higgs particle would contribute to electroweak 
observables only at two-loop and higher orders, and is therefore practically 
unconstrained by precision electroweak measurements \cite{vanderBij:1985ww}.

\vspace*{0.3cm}
\renewcommand{\thesection}{\arabic{section}}
\section{~Recent discovery of a Higgs-boson candidate at LHC}
\vspace*{0.3cm}

\begin{figure}[h!]
\begin{center}
\epsfig{file=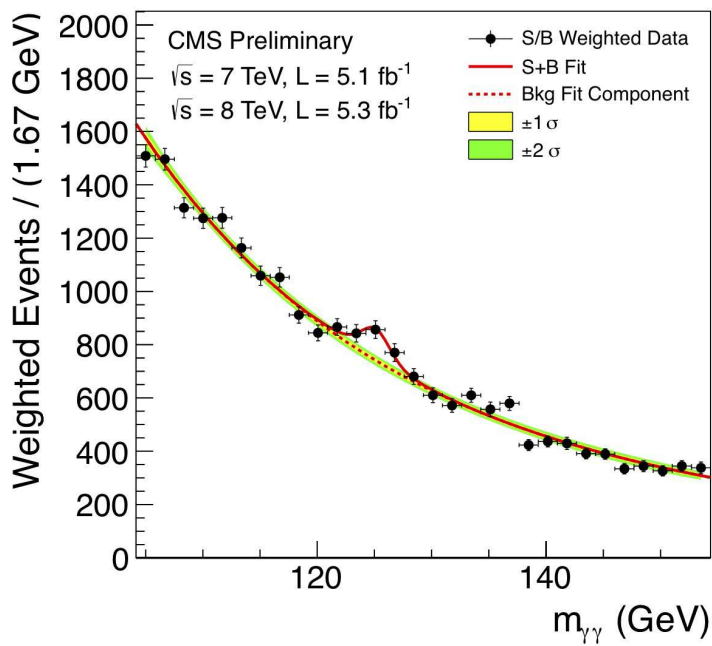,height=0.35\textheight}
\end{center}
\vskip -2mm
\caption{Di-photon ($\gamma\gamma$) invariant mass distribution for the CMS
data up to June 2012 (black points with error bars). The solid red line shows
the fit result for signal plus background; the dashed red line shows only the
background. This distribution and a similar plot by the ATLAS collaboration
were first presented in July 2012.}
\label{fig:CMS}
\end{figure}

~~~~Preliminary results on searches for a SM Higgs boson were presented in
July 2012 by the ATLAS and CMS collaborations at CERN's Large Hadron Collider
(LHC). A state decaying to several distinct final states (`channels') has
been observed with a statistical significance of five standard deviations
(5 $\sigma$). The evidence is strongest in the two channels with the best mass
resolution: the two-photon channel (see Fig.\,\ref{fig:CMS}) and the final 
state with two pairs of charged leptons (electrons or muons). 

The observed state has a mass of about 126 GeV. Its production rate is
consistent, within the present statistical and systematic uncertainties, with
the predicted rate for the SM Higgs boson. Event yields in different
production topologies and different decay modes are self-consistent.
  
The CMS data rule out, at 95\% confidence level (CL), the existence of the SM 
Higgs boson in the two broad mass ranges of 110--122.5 GeV and 127--600 GeV;
the ATLAS data exclude it at 99\% CL in the mass region 110--600 GeV,
except in the narrow range 121.8 to 130.7 GeV. Masses up to $\sim 115$ GeV were
already excluded by CERN's LEP collider at a similar confidence level.

\vspace*{0.3cm}
\renewcommand{\thesection}{\arabic{section}}
\section{~Single Higgs production in $\gamma\gamma$ collisions}
\vspace*{0.3cm}

~~~~Since photons couple directly to all fundamental fields carrying the
electromagnetic current (leptons, quarks, W bosons, supersymmetric particles),
$\gamma\gamma$ collisions provide a comprehensive means of exploring virtually
every aspect of the SM and its extensions (see \cite{boos, belusev} and
references therein). Moreover, the cross-sections for production of 
charged-particle pairs in $\gamma\gamma$ interactions are approximately an order
of magnitude larger than in $e^{+}e^{-}$ annihilations. 

In $\gamma\gamma$ collisions, the {\em Higgs boson} is produced as a 
single resonance in a state of definite CP, which is perhaps the most important
advantage over $e^{+}e^{-}$ annihilations, where this $s$-channel process is
highly suppressed. For the Higgs mass in the range $\mbox{\large{$m$}}_{\mbox
{\tiny{H}}}^{~} =$ 115$-$200 GeV, the effective cross-section for $\gamma\gamma
\rightarrow$ H is about a factor of five larger than that for Higgs production
in $e^{+}e^{-}$ annihilations. In this mass range, the process $e^{+}e^{-}
\rightarrow {\rm HZ}$ requires considerably higher center-of-mass energies than
$\gamma\gamma \rightarrow$ H. 

In $e^{+}e^{-}$ annihilations, the heavy neutral MSSM Higgs bosons can be
created only by associated production ($e^{+}e^{-} \rightarrow H^{0}A^{0}$), 
whereas in $\gamma\gamma$ collisions they are produced as single resonances 
($\gamma\gamma \rightarrow H^{0},\,A^{0}$) with masses up to 80\% of the initial
$e^{-}e^{-}$ collider energy \cite{zerwas}. For example, if $H^{0}$ and $A^{0}$
have the same mass of about 500 GeV, then they could be produced either in pairs
in $e^{+}e^{-}$ annihilations at CM energies ${\rm E}_{ee} \sim 1$ TeV, or as 
single particles in $\gamma\gamma$ collisions at ${\rm E}_{ee} \sim 600$ GeV. 
 
The reaction $\gamma\gamma \rightarrow$ H, which is related to ${\rm H}
\rightarrow \gamma\gamma$, proceeds through a `loop diagram' and receives
contributions from {\em all} charged particles that couple to the photon and
the Higgs boson. Thus, the {\em two-photon width} $\Gamma ({\rm H}
\rightarrow \gamma\gamma )$ is sensitive to the Higgs-top Yukawa coupling, as
well as mass scales far beyond the energy of the $\gamma\gamma$ collision.
Assuming that the branching ratio ${\rm BR}({\rm H} \rightarrow b\bar{b})$ can
be measured to an accuracy of about 2\% in the process $e^{+}e^{-} \rightarrow
{\rm HZ}$, the $\gamma\gamma$ partial width can be determined with a similar 
precision for $\mbox{\large{$m$}}_{\mbox{\tiny{H}}}^{~} \simeq 120$ GeV by
measuring the cross-section $\mbox{\large{$\sigma$}}(\gamma\gamma \rightarrow
{\rm H} \rightarrow b\bar{b}) \propto \Gamma ({\rm H} \rightarrow \gamma\gamma)
{\rm BR}({\rm H} \rightarrow b\bar{b})$. 

\begin{figure}[t]
\begin{center}
\epsfig{file=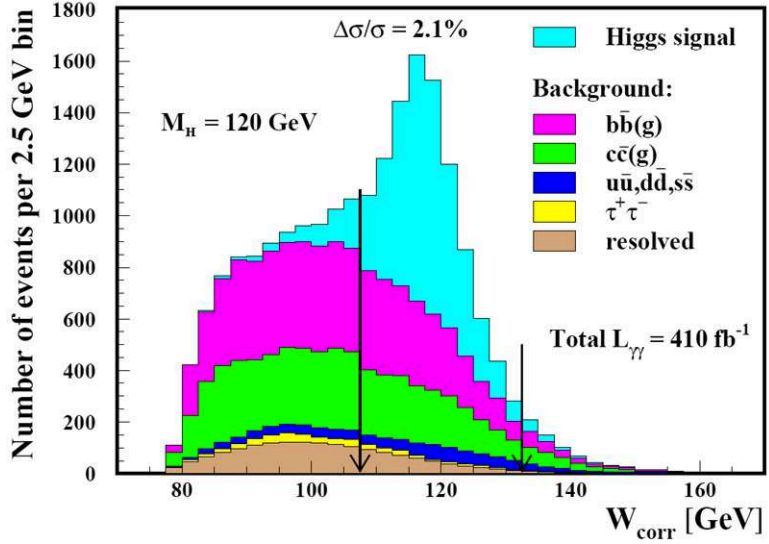,height=0.30\textheight}
\end{center}
\vskip -6mm
\caption{The reconstructed invariant-mass distribution of the $\gamma\gamma
\rightarrow {\rm H} \rightarrow b\bar{b}$ signal and the $b\bar{b}(g)$ and
$c\bar{c}(g)$ backgrounds. The gluon (`resolved') structure of the photon can 
be measured {\em in situ}. Credit: P. Niezurawski, A. Zarnecki and
M. Krawczyk.}
\label{fig:invariant_mass}
\end{figure}

High-energy photons can be produced by Compton-backscattering of laser light on
electron beams. Both the energy spectrum and polarization of the backscattered 
photons depend strongly on the polarizations of the incident electrons and
laser photons. The key advantage of using $e^{-}e^{-}$ beams is that they can
be polarized to a high degree, enabling one to tailor the photon energy 
distribution to one's needs. In a collision of two photons, the possible 
helicities are 0 or 2. For example, the Higgs boson is produced in the 
$J_{z} = 0$ state, whereas the background processes $\gamma\gamma \rightarrow 
b\bar{b},\,c\bar{c}$ are suppressed for this helicity configuration
(see Fig.\,\ref{fig:invariant_mass}). The circular polarization of the photon 
beams is therefore an important asset, for it can be used both to enhance the 
signal and suppress the background. 

The {\em CP properties} of any neutral Higgs boson that may be produced 
at a photon collider can be {\em directly} determined by controlling the 
polarizations of Compton-scattered photons \cite{grzadkowski}. A CP-even Higgs
boson couples to the combination ${\bf e}_{\mbox{\tiny{1}}}\mbox{\boldmath
{$\cdot$}}\,{\bf e}_{\mbox{\tiny{2}}}$, whereas a CP-odd Higgs boson couples to
$({\bf e}_{\mbox{\tiny{1}}}\!\times\!{\bf e}_{\mbox{\tiny{2}}})\,\mbox{\boldmath
{$\cdot$}}\,\mbox{\boldmath{$k$}}_{\gamma}$:
\begin{eqnarray*}
{\cal M}(\gamma\gamma\rightarrow{\rm H}[0^{++}]) \!&\propto&\!{\bf e}_{\mbox
{\tiny{1}}}\mbox{\boldmath{$\cdot$}}\,{\bf e}_{\mbox{\tiny{2}}} \,\propto\,
1 + \cos 2\phi \\*[1mm]
{\cal M}(\gamma\gamma\rightarrow A[0^{-+}]) \!&\propto&\!({\bf e}_{\mbox
{\tiny{1}}}\!\times\!{\bf e}_{\mbox{\tiny{2}}})\,\mbox{\boldmath{$\cdot$}}\,
\mbox{\boldmath{$k$}}_{\gamma} \,\propto\, 1 - \cos 2\phi
\end{eqnarray*}
where ${\bf e}_{i}^{~}$ are polarization vectors of colliding photons, $\phi$ is
the angle between them, and $\mbox{\boldmath{$k$}}_{\gamma}$ is the momentum
vector of one of the Compton-scattered photons; $0^{++}$ and $0^{-+}$ are the
quantum numbers ${\rm J}^{\rm PC}$. The scalar (pseudoscalar) Higgs boson 
couples to {\em linearly polarized} photons with a maximum strength if the 
polarization vectors are parallel (perpendicular): $\mbox{\large{$\sigma$}}
 \propto 1 \pm l_{\mbox{\tiny{1}}}l_{\mbox{\tiny{2}}}\cos 2\phi$, where $l_{i}$
are the degrees of linear polarization; the signs $\pm$ correspond to the
${\rm CP} = \pm 1$ particles. 

The general amplitude for a CP-{\em mixed state} to couple to two photons is
\begin{equation}
{\cal M} \,=\, {\cal E}({\bf e}_{\mbox{\tiny{1}}}\mbox{\boldmath{$\cdot$}}\,
{\bf e}_{\mbox{\tiny{2}}}) \,+\, {\cal O}({\bf e}_{\mbox{\tiny{1}}}\!\times\!
{\bf e}_{\mbox{\tiny{2}}})_{z}^{~}
\end{equation}
where ${\cal E}$ is the CP-even and ${\cal O}$ the CP-odd contribution to the 
amplitude. If we denote the {\em helicities} of the two photons by $\lambda_{1}$
and $\lambda_{2}$, with $\lambda_{1},\lambda_{2} = \pm 1$, then the above
vector products can be expressed as 
\[ {\bf e}_{\mbox{\tiny{1}}}\mbox{\boldmath{$\cdot$}}\,{\bf e}_{\mbox{\tiny{2}}}
 \,=\, -(1 + \lambda_{1}\lambda_{2})/2\hspace*{1cm}\mbox{and}\hspace*{1cm}
({\bf e}_{\mbox{\tiny{1}}}\!\times\!{\bf e}_{\mbox{\tiny{2}}})_{z}^{~} \,=\,
 i\lambda_{1}(1 + \lambda_{1}\lambda_{2})/2 \]
Now, 
\[|{\cal M}_{++}|^{2} - |{\cal M}_{--}|^{2} = -4{\rm Im}({\cal E}{\cal O}^{*})\]
\[2{\rm Re}({\cal M}_{--}^{*}{\cal M}_{++})=2(|{\cal E}|^{2} -|{\cal O}|^{2})\]
\[2{\rm Im}({\cal M}_{--}^{*}{\cal M}_{++}) = -4{\rm Re}({\cal E}{\cal O}^{*})\]
When these expressions are divided by 
\[ |{\cal M}_{++}|^{2} + |{\cal M}_{--}|^{2} \,=\, 2(|{\cal E}|^{2} + 
|{\cal O}|^{2}) \]
one obtains three {\em polarization asymmetries} that yield an unambiguous 
measure of CP-mixing \cite{grzadkowski}. It is necessary to have both 
{\em linearly} and {\em circularly} polarized photons in order to measure those
asymmetries. 

In $e^{+}e^{-}$ annihilations, it is possible to discriminate between
CP-even and CP-odd neutral Higgs bosons, but would be difficult to detect small
CP-violating effects (which contribute only at the one-loop level) for a
dominantly CP-even component (which contributes at the tree level in 
$e^{+}e^{-}$ collisions) \cite{hagiwara}.

\vspace*{0.3cm}
\section{~Single Higgs production in $e^{+}e^{-}$ annihilations}
\vspace*{0.3cm}

\begin{figure}[h] 
\begin{center}
\epsfig{file=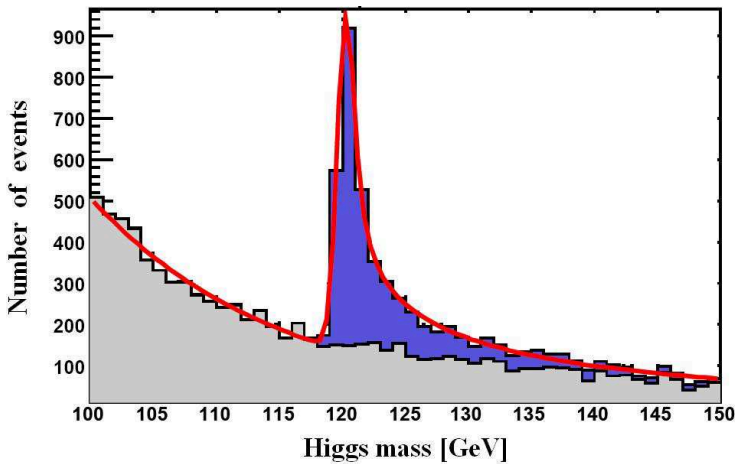,height=0.25\textheight}
\end{center}
\vskip -6mm
\caption{Distribution of the invariant mass of the system recoiling against a
pair of leptons in the process $e^{+}e^{-} \rightarrow {\rm HZ} \rightarrow
X\ell^{+}\ell^{-}$ for $\mbox{\large{$m$}}_{\mbox{\tiny{H}}}^{~} = 120$ GeV and
$\int\!{\cal L} = 500\,fb^{-1}$ at $\sqrt{s} = 250$ GeV. The red line is a fit 
to a Monte Carlo simulation of the Higgs signal and the ZZ background; the gray
area represents the background only \cite{Abe}. For $\mbox{\large{$m$}}_{\mbox
{\tiny{H}}}^{~} \simeq 120$ GeV, the optimum center-of-mass energy is $\sqrt{s}
\simeq 230$ GeV.}
\label{fig:recoil_mass}
\end{figure}

~~~~A particularly noteworthy feature of an $e^{+}e^{-}$ collider is that
the Higgs boson can be detected in the {\em Higgs-strahlung process}
\begin{equation}
e^{+}e^{-} \,\rightarrow\, {\rm HZ}
\end{equation}
even if it decays into invisible particles (e.g., the lightest {\em neutralino}
of a supersymmetric model). In this case the signal manifests itself as a peak
in the distribution of invariant mass of the system recoiling against the 
lepton pair stemming from Z-boson decay (see Fig.\,\ref{fig:recoil_mass}).

By exploiting the ${\rm HZ} \rightarrow X\ell^{+}\ell^{-}$ channel, the
Higgs-strahlung {\em cross-sections} can be measured with statistical
errors of 2.6 to 3.1 percent for Higgs-boson masses from 120 to 160 GeV (see
\cite{heinemeyer} and references therein).

From the fits to the reconstructed mass spectra in the channels ${\rm HZ} 
\rightarrow q\bar{q}\ell^{+}\ell^{-},~b\bar{b}q\bar{q},~{\rm WW}\ell^{+}
\ell^{-}$ and ${\rm WW}q\bar{q}$, the {\em Higgs-boson mass} can be 
determined with an uncertainty of 40 to 70 MeV for $\mbox{\large{$m$}}_{\mbox
{\tiny{H}}}^{~}$ in the range 120 to 180 GeV \cite{heinemeyer}.

To determine the {\em spin} and {\em parity} of the SM Higgs boson
in the Higgs-strahlung process, one can use the information on (1) the energy
dependence of the Higgs-boson production cross-section just above the kinematic
threshold, and (2) the angular distribution of the Z/H bosons. The best way to
study the {\em CP properties} of the Higgs boson is by analyzing the spin
correlation effects in the decay channel ${\rm H} \rightarrow \tau^{+}\tau^{-}$
(see \cite{heinemeyer} and references therein). 
 
The Higgs-strahlung cross-section, which dominates at low CM energies,
decreases with energy in proportion to $1/s$ (see Eq. (20)). In contrast, the 
cross-section for the {\em W-fusion process}
\begin{equation}
e^{+}e^{-} \,\rightarrow\, {\rm H}\nu_{e}^{~}\bar{\nu}_{e}^{~}
\end{equation}
increases with energy in proportion to log($s/\mbox{\large{$m$}}_{\mbox{\tiny
{H}}}^{\,2}$), and hence becomes more important at energies $\sqrt{s} \gsim 
500$ GeV for the Higgs-mass range $115~{\rm GeV} \lsim \mbox{\large{$m$}}_
{\mbox{\tiny{H}}}^{~} \lsim 200~{\rm GeV}$.

\vspace*{0.3cm}
\section{~Higgs-pair production in $\gamma\gamma$ and $e^+e^-$ collisions}
\vspace*{0.3cm}

~~~~It is well known that hadron colliders are not ideally suited for measuring
the self-coupling of the Higgs boson if $\mbox{\large{$m$}}_{\mbox{\tiny{H}}}
^{~}\,\leq\,140$~GeV \cite{baur}. The potential of a future $\gamma\gamma /
e^+e^-$ collider for determining the HHH coupling has therefore been closely 
examined (see \cite{belusev1} and [16--20]).

\begin{figure}[t]
\vskip -3mm
\setlength{\unitlength}{1cm}
\begin{picture}(10,10)
\put(0.25,0){\epsfig{file=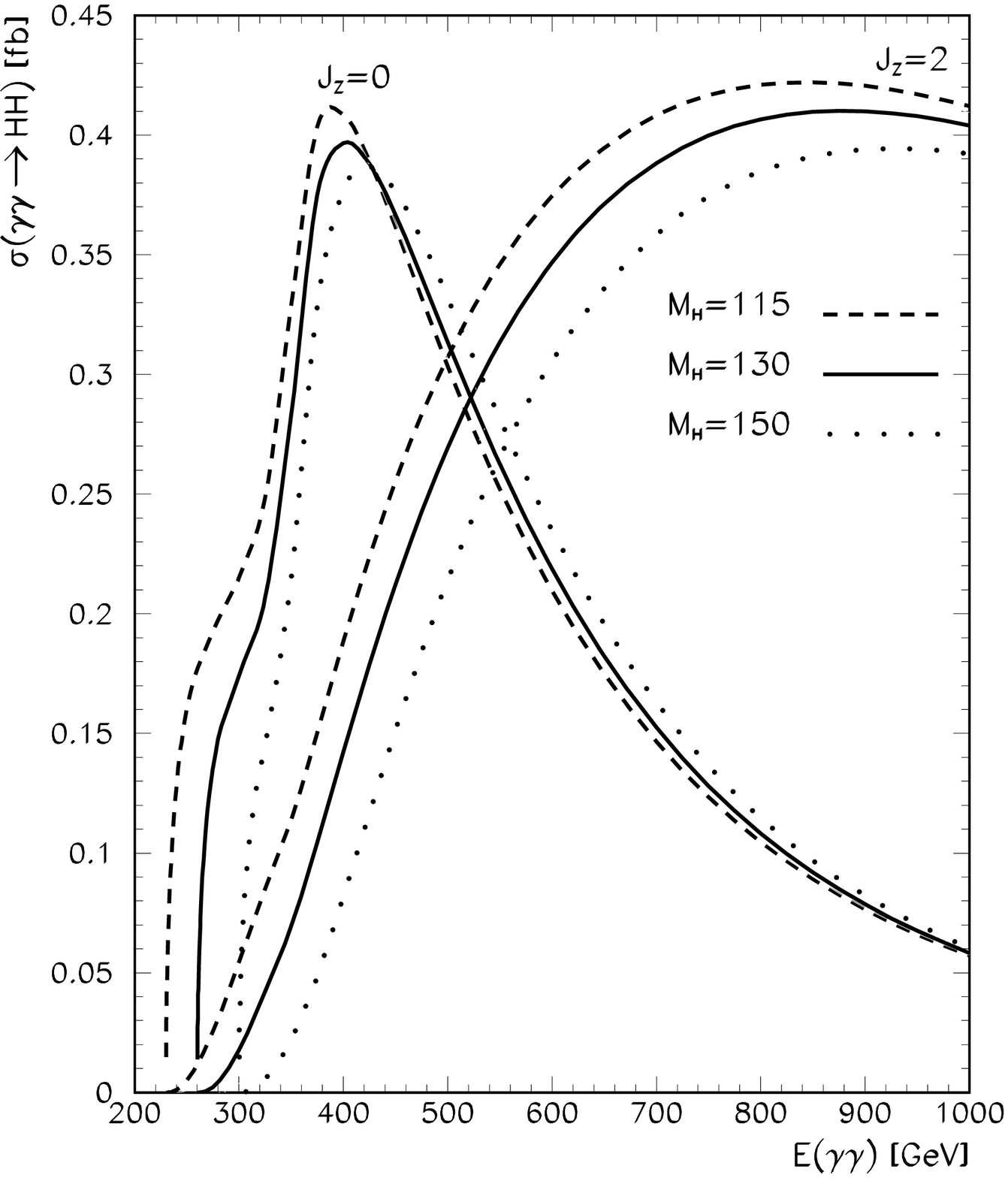,width=0.5\textwidth}}
\put(8.5,0){\epsfig{file=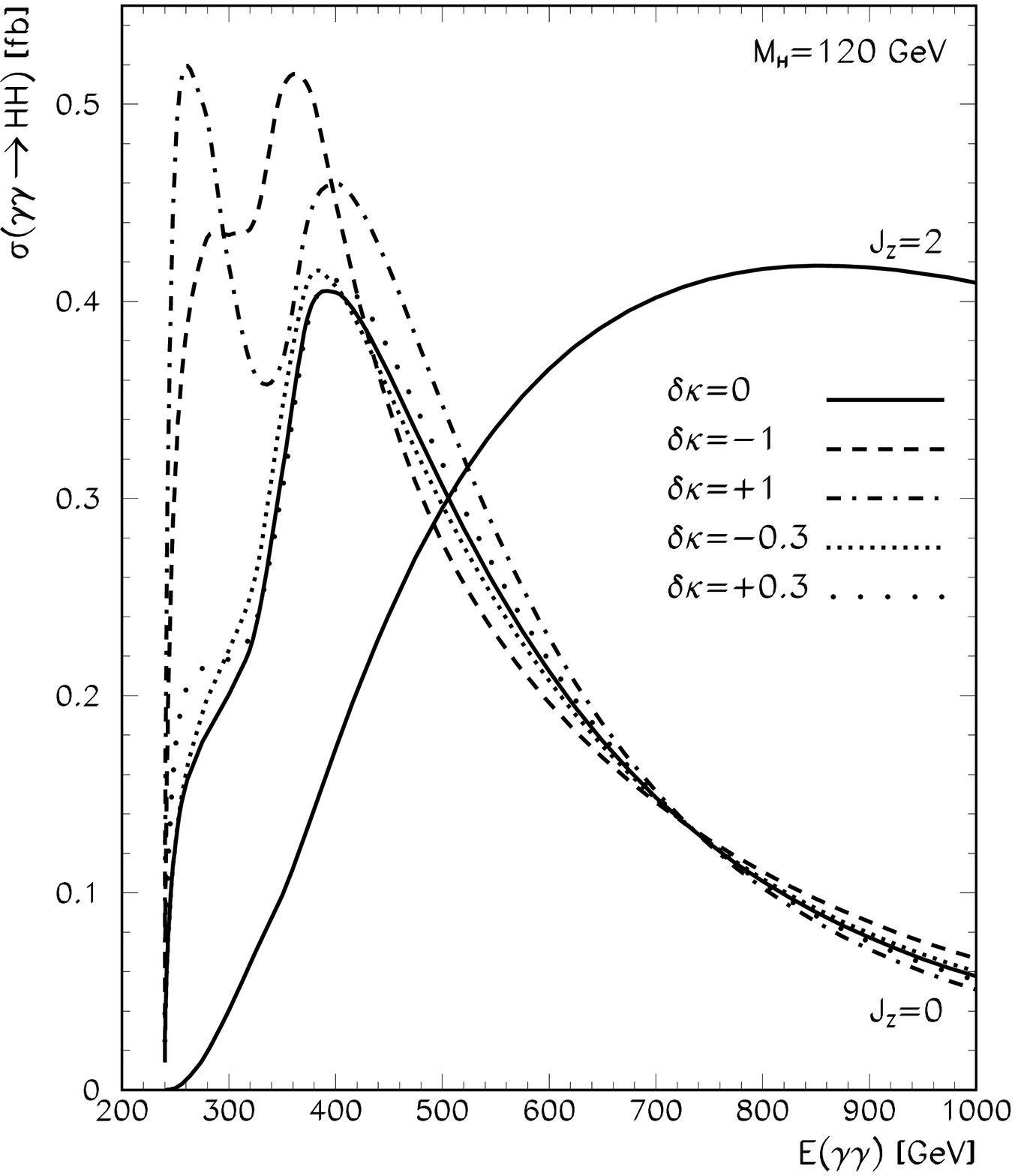,width=0.5\textwidth}}
\end{picture}
\vskip -8mm
\caption{(a) The total $\gamma\gamma\to{\rm HH}$ cross-section as a
  function of the $\gamma\gamma$ center-of-mass energy. Contributions
  for equal ($J_z=0$) and opposite ($J_z=2$) photon helicities are
  shown separately.
  \newline
  (b) The cross-sections for HH production in $\gamma\gamma$ collisions for
$\mbox{\large{$m$}}_{\mbox{\tiny{H}}}^{~} = 120$ GeV and anomalous trilinear 
Higgs self-couplings $\delta\kappa=0,\pm 1, \pm 0.3$; Credit: R.Belusevic and
G. Jikia \cite{belusev1}.}
\label{fig:gghh}
\end{figure}

The production of a pair of SM Higgs bosons in photon-photon collisions,
\begin{equation}
\gamma\gamma \,\to\, {\rm HH}
\end{equation}
which is related to the Higgs-boson decay into two
photons, is due to W-boson and top-quark box and triangle loop diagrams. The
total cross-section for $\gamma\gamma\to{\rm HH}$ in polarized photon-photon
collisions, calculated at the leading one-loop order \cite{Jikia:1992mt} as a
function of the $\gamma\gamma$ center-of-mass energy and for 
$\mbox{\large{$m$}}_{\mbox{\tiny{H}}}^{~}$
between 115 and 150 GeV, is shown in Fig.\,\ref{fig:gghh}a. The
cross-section calculated for equal photon helicities, $\mbox{\large{$\sigma$}}_
{\gamma\gamma\,\rightarrow\,\mbox{\tiny{HH}}}(\mbox{\small{$J_{z}=0$}})$, rises
sharply above the $2\mbox{\large{$m$}}_{\mbox{\tiny{H}}}^{~}$ 
threshold for different values of $\mbox{\large{$m$}}_{\mbox{\tiny{H}}}^{~}$,
and has a peak value of about $0.4$~fb at a 
$\gamma\gamma$ center-of-mass energy of 400~GeV. In contrast, the cross-section
for opposite photon helicities, $\mbox{\large{$\sigma$}}_{\gamma\gamma\,
\rightarrow\,\mbox{\tiny{HH}}}(\mbox{\small{$J_{z}=2$}})$, rises more slowly
with energy because a pair of Higgs bosons is produced in a state with orbital
angular momentum of at least $2\hbar$.

The cross-sections for equal photon helicities are of special interest, since
only the $J_z=0$ amplitudes contain contributions with trilinear Higgs 
self-coupling. By adding to the SM Higgs potential $V(\Phi^{\dag}\Phi )$ 
a gauge-invariant dimension-6 operator $\mbox{\large{$($}}\Phi^{\dag}\Phi\mbox
{\large{$)$}}^{3}$, one introduces a gauge-invariant anomalous trilinear
Higgs coupling $\delta\kappa$ \cite{Jikia:1992mt}. For the reaction $\gamma
\gamma \to {\rm HH}$, the only effect of such a coupling in the {\em unitary
gauge} would be to replace the trilinear HHH coupling of the SM in Eq.
\ref{eq:hhh} by an {\em anomalous Higgs self-coupling} 
\begin{equation}
\lambda \,=\, (1 + \delta\kappa )\lambda_{\mbox{\tiny{HHH}}}^{~} 
\end{equation}
The dimensionless anomalous coupling $\delta\kappa$ is normalized so that 
$\delta\kappa=-1$ exactly cancels the SM HHH coupling. The cross-sections
$\mbox{\large{$\sigma$}}_{\gamma\gamma\,\rightarrow\,\mbox{\tiny{HH}}}$ for
five values of $\delta\kappa$ are shown in Fig.~\ref{fig:gghh}b.

In an experiment to measure the trilinear Higgs self-coupling, the contribution
from $\gamma\gamma \to {\rm HH}$ for opposite photon helicities represents
an irreducible background. However, this background is suppressed if one
chooses a $\gamma\gamma$ center-of-mass energy below about 320 GeV. 

To ascertain the potential of $\gamma\gamma$ colliders for measuring an 
anomalous trilinear Higgs self-coupling, one must take into account the fact 
that the photons are not monochromatic \cite{Ginzburg:1982yr}.
It is envisaged that an $e^{-}e^{-}$ linac and a terawatt laser system will be
used to produce Compton-scattered $\gamma$-ray beams for a photon collider.
Both the energy spectrum and polarization of the backscattered photons depend
strongly on the polarizations of the incident electrons and photons. A
longitudinal electron-beam polarization of 90\% and a 100\% circular
polarization of laser photons are assumed throughout.

\begin{figure}[t]
\vskip -3mm
\setlength{\unitlength}{1cm}
\begin{picture}(10,10)
\put(0.25,0){\epsfig{file=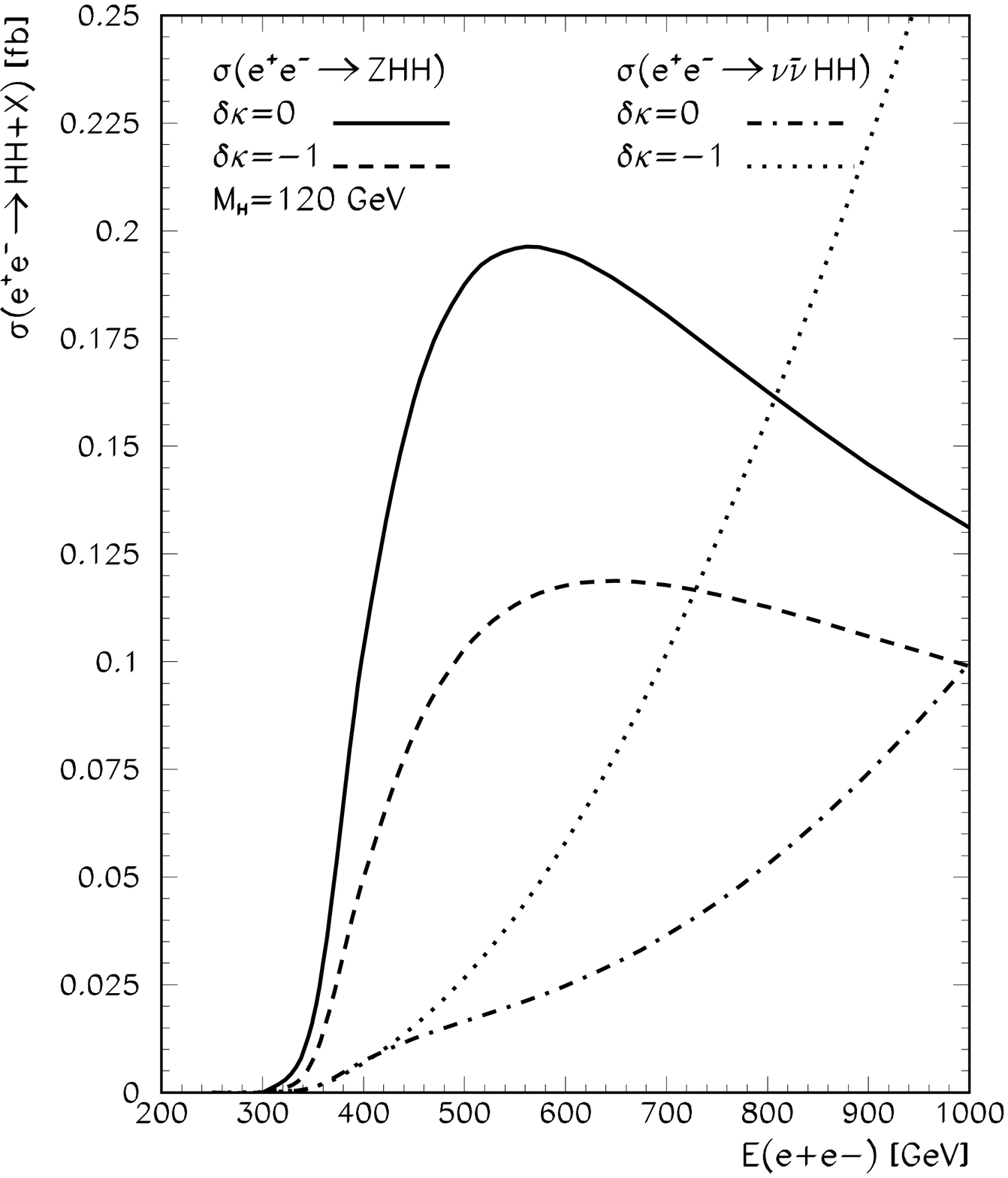,width=0.5\textwidth}}
\put(8.5,0){\epsfig{file=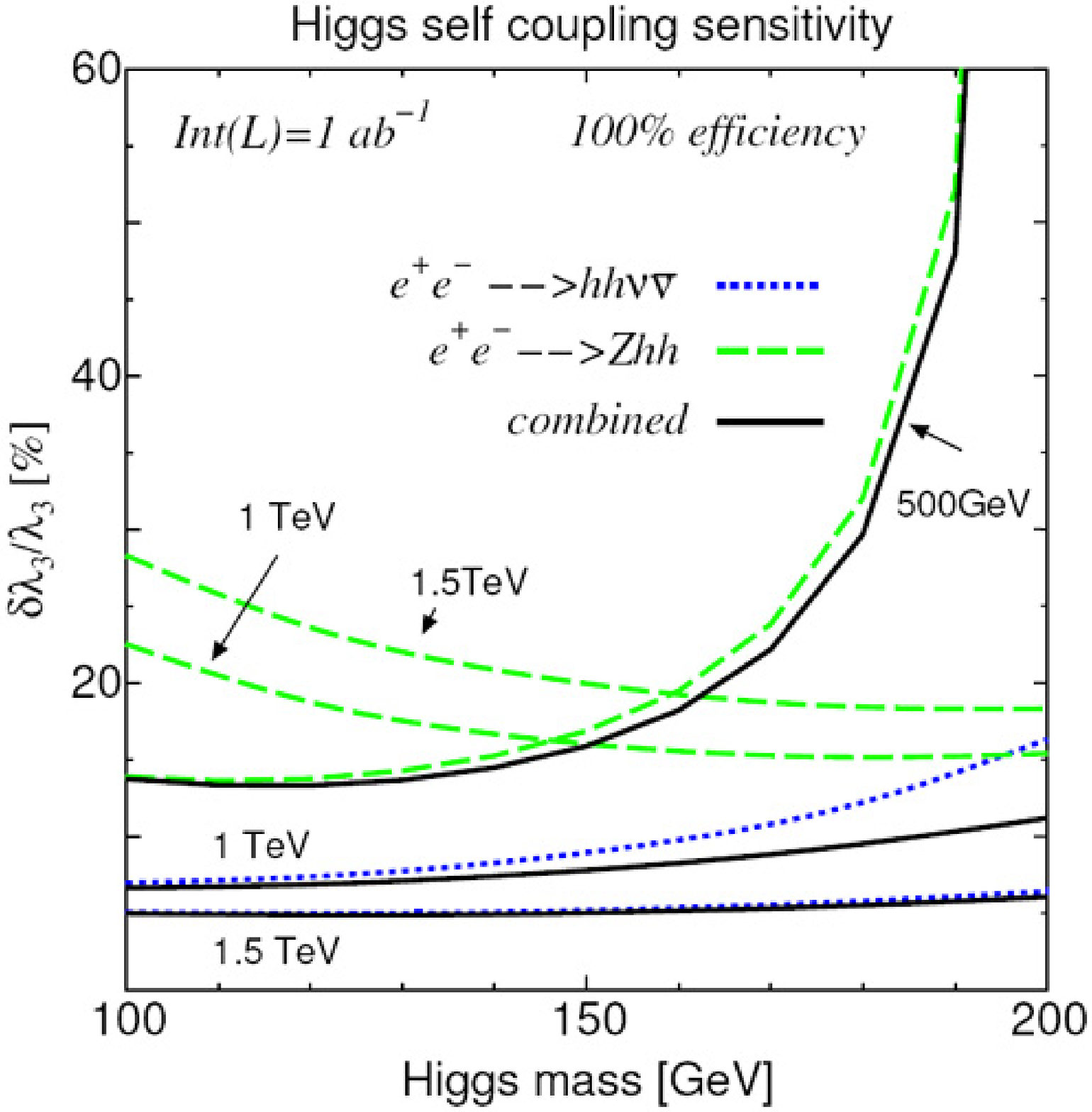,width=0.48\textwidth}}
\end{picture}
\vskip -1mm
\caption{(a) Total cross-sections for $e^+e^-\,\to\, {\rm HH}Z$ and 
$e^+e^-\,\to\,{\rm HH}\nu_e\bar\nu_e$ as functions of the $e^+e^-$
CM energy for $\mbox{\large{$m$}}_{\mbox{\tiny{H}}}^{~}=120$
GeV and the anomalous trilinear Higgs self-couplings $\delta\kappa=0$ or $-1$
\cite{belusev1}. (b) Statistical sensitivity of the trilinear self-coupling for
the processes $e^+e^-\,\to\,{\rm HH}Z$ and $e^+e^-\,\to\,{\rm HH}\nu_e\bar\nu_e$
\cite{GLC}.} 
\label{fig:eehh}
\end{figure}

The trilinear self-coupling of the Higgs boson can also be measured either in 
the so-called {\em double Higgs-strahlung process}
\begin{equation}
e^+e^- \,\to\, {\rm HHZ}
\end{equation}
or in the {\em W-fusion reaction}
\begin{equation}
e^+e^- \,\to\, {\rm HH}\nu_e^{~}\bar{\nu}_e^{~}
\end{equation}
The total cross-section for pair production of 120-GeV Higgs bosons in $e^+e^-$
collisions, calculated for {\em unpolarized} beams, is presented in
Fig.~\ref{fig:eehh} for anomalous trilinear Higgs self-couplings $\delta
\kappa=0$ or $-1$. If the electron beam is 100\% polarized, the double
Higgs-strahlung cross-section will stay approximately the same, while the
W-fusion cross-section will be twice as large. From Fig.~\ref{fig:eehh}
we infer that the SM double Higgs-strahlung cross-section exceeds 0.1~fb at 
400~GeV for $\mbox{\large{$m$}}_{\mbox{\tiny{H}}}^{~}\,=\,120$~GeV, and 
reaches a broad maximum of about 0.2~fb at a CM energy of 550~GeV. The SM 
cross-section for W-fusion stays below 0.1~fb for CM energies up to 1 TeV.  

For $\mbox{\large{$m$}}_{\mbox{\tiny{H}}}^{~}\,=\,120$~GeV, and assuming a
longitudinal electron-beam polarization of 90\%, the maximum sensitivity to an
anomalous trilinear Higgs self-coupling is achieved in the double 
Higgsstrahlung process at a CM energy of about 500~GeV \cite{belusev1}. This is
significantly higher than the optimal CM energy in $\gamma\gamma$ collisions.
In the W-fusion process, a similar sensitivity is attained at CM energies well 
above 500 GeV.

Calculations show that the {\em statistical} sensitivity of
$\mbox{\large{$\sigma$}}_{\gamma\gamma\,\rightarrow\,\mbox{\tiny{HH}}}$ to the
Higgs self-coupling is maximal near the kinematic threshold for Higgs-pair
production if $\mbox{\large{$m$}}_{\mbox{\tiny{H}}} \sim 120$ GeV,
and is comparable with the sensitivities of $\mbox{\large{$\sigma$}}_{e^{+}
e^{-}\,\rightarrow\,\mbox{\tiny{HHZ}}}$ and $\mbox{\large{$\sigma$}}_{e^{+}
e^{-}\,\rightarrow\,\mbox{\tiny{HH}}\nu\bar{\nu}}$ to this coupling for 
${\rm E}_{ee}\leq 700$ GeV, even if the integrated luminosity in $\gamma\gamma$
collisions is only one third of that in $e^+e^-$ annihilations \cite{belusev1}.
The overall {\em acceptance} should, in principle, be considerably larger in the
process $\gamma\gamma \rightarrow {\rm HH}$ than in the reaction $e^{+}e^{-} 
\rightarrow {\rm HHZ}$ due to the smaller final-state particle multiplicity.

The Feynman diagrams for the process $\gamma\gamma \rightarrow {\rm HH}$ are
shown in Fig.~1 of \cite{Jikia:1992mt}. New physics beyond the Standard Model
introduces additional complexity into the subtle interplay between the
Higgs `pole amplitudes' and the top-quark and W-boson `box diagrams':
\[ |{\cal M}(J_{z} = 0)|^{2} \,=\, |A(s)(\lambda_{\mbox{\tiny{SM}}}^{~} +
\delta\lambda ) \,+\, B|^{2} \]
where $\lambda_{\mbox{\tiny{SM}}}^{~}$ is the trilinear Higgs self-coupling in
the SM. From the above expression we infer that the cross-section for 
$\gamma\gamma \rightarrow {\rm HH}$ is a quadratic function of
$\lambda \equiv \lambda_{\mbox{\tiny{SM}}}^{~} + \delta\lambda$:
\[ \mbox{\large{$\sigma$}}(\lambda) \,=\, \alpha\lambda^{2} + \beta\lambda + 
\gamma~~~~~~~~~~~~~~~\alpha > 0,~~\gamma > 0 \]

There are various ways to define the sensitivity of the trilinear Higgs
self-coupling. For instance, we can expand around $\mbox{\large{$\sigma$}} =
\mbox{\large{$\sigma$}}_{\mbox{\tiny{SM}}}^{~}$, and express the number of
events as
\[ N \,=\, L\,\mbox{\large{$\sigma$}}_{\mbox{\tiny{SM}}}^{~} \,+\, L\,\delta
\lambda\!\left (\frac{\raisebox{-.4ex}{${\rm d}\mbox{\large{$\sigma$}}$}}
{{\rm d}\lambda}\right )_{\!\lambda\,=\,\lambda_{\mbox{\tiny{SM}}}^{~}} \,+\,
\cdots \]
where $L$ is the integrated luminosity. The sensitivity of $\lambda$ is given
by
\[ \sqrt{N} \,=\, \left |L\,\delta\lambda\!\left (\frac{\raisebox{-.4ex}
{${\rm d}\mbox{\large{$\sigma$}}$}}{{\rm d}\lambda}\right )_
{\!\lambda\,=\,\lambda_{\mbox{\tiny{SM}}}^{~}}\right | \]
i.e.,
\[ \delta\lambda \,=\, \frac{\sqrt{L\,\mbox{\large{$\sigma$}}_{\mbox{\tiny
{SM}}}^{~}}}{L\,({\rm d}\mbox{\large{$\sigma$}}/{\rm d}\lambda )_{\lambda\,=\,
\lambda_{\mbox{\tiny{SM}}}^{~}}^{~}} \,=\, \frac{\sqrt{\mbox{\large{$\sigma$}}_
{\mbox{\tiny{SM}}}^{~}/L}}{({\rm d}\mbox{\large{$\sigma$}}/{\rm d}\lambda )_
{\lambda\,=\,\lambda_{\mbox{\tiny{SM}}}^{~}}^{~}} \] 

\begin{figure}[t]
\begin{center}
\epsfig{file=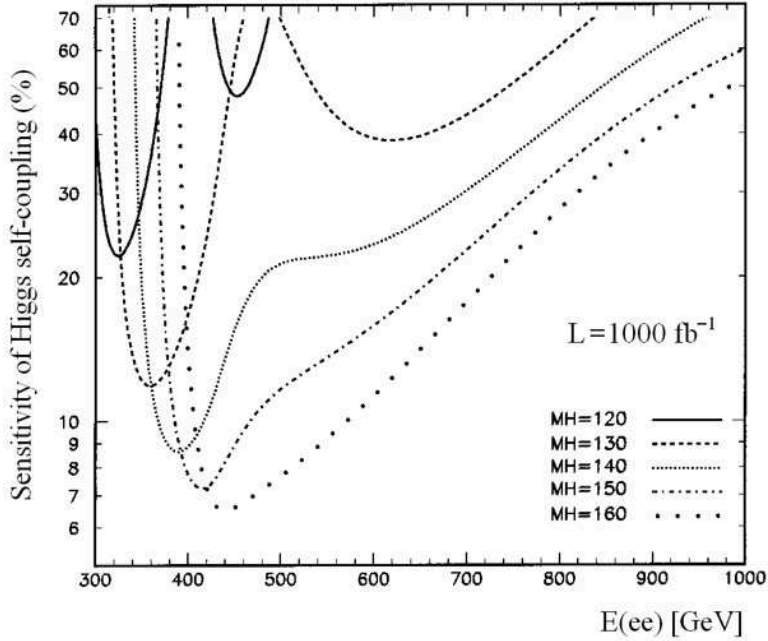,height=0.35\textheight}
\end{center}
\vskip -6mm
\caption{Statistical sensitivity of the trilinear Higgs self-coupling for
various Higgs-boson masses assuming $\int\!{\cal L} = 1000\,fb^{-1}$. Based on
the calculation by R. Belusevic and G. Jikia \cite{belusev1}.}
\label{fig:sensitivity}
\end{figure}

A plot of the trilinear Higgs self-coupling sensitivity in $\gamma\gamma$
collisions, based on the above expression for $\delta\lambda$, is shown in
Fig.~\ref{fig:sensitivity}; for $e^{+}e^{-}$ annihilations, see Fig.~3.8 in
\cite{heinemeyer}. An obvious drawback of the above definition of 
$\delta\lambda$ is that its value becomes unphysically large when the 
derivative ${\rm d}\mbox{\large{$\sigma$}}/{\rm d}\lambda \rightarrow 0$, which
means that one should take into account also the $\lambda^{2}$ term. 

Since the cross-section
$\mbox{\large{$\sigma$}}_{\gamma\gamma\,\rightarrow\,\mbox{\tiny{HH}}}$ does
not exceed 0.4 fb, it is essential to attain the highest possible luminosity,
rather than energy, in order to measure the trilinear Higgs self-coupling.
As shown in \cite{belusev1}, 
appropriate angular and invariant-mass cuts and a $b$-tagging
requirement, which result in a Higgs-pair reconstruction efficiency of about
50\%, would suppress the dominant W-pair and four-quark backgrounds well
below the HH signal. For such a reconstruction efficiency, a center-of-mass
energy E$_{ee}\approx 300$~GeV and $\mbox{\large{$m$}}_{\mbox{\tiny{H}}} = 120$
GeV an integrated $\gamma\gamma$
luminosity $L_{\gamma\gamma} \approx 450$~fb$^{-1}$ would be needed to exclude
a zero trilinear Higgs-boson self-coupling at the 5$\sigma$ level (statistical
uncertainty only). An even higher luminosity is required for an accurate
measurement of this coupling.

The results of detailed feasibility studies for measuring Higgs-pair production 
in $\gamma\gamma$ and $e^+e^-$ collisions have been reported 
\cite{kawada, tian}. It has been shown that the optimum $\gamma\gamma$ 
collision energy is around 270 GeV for a 120-GeV Higgs boson, and that the main
backrounds at this energy are the processes $\gamma\gamma \rightarrow$ WW, ZZ 
and $b\bar{b}b\bar{b}$. The preliminary analysis described in \cite{kawada}
suggests that $\gamma\gamma \rightarrow$ HH could be observed with a statistical
significance of $\sim 5 \sigma$ provided proper color-singlet clustering is 
used in jet reconstruction.  

\vspace*{0.3cm}
\section{~Higgs couplings to SM particles}
\vspace*{0.3cm}

~~~~In the {\em unitary gauge}, the kinetic term in Eq. (1) can be expressed as
\begin{equation}
(D_{\mu}\Phi )^{\dag}(D^{\mu}\Phi ) \,=\, \frac{\raisebox{-.3ex}{\mbox{\small
{1}}}}{\raisebox{.2ex}{\mbox{\small{2}}}}\mbox{\large{$($}}\partial_{\mu\,}
{\rm H}\mbox{\large{$)$}}^{2} \,+\, \frac{g^{2}}{\raisebox{.2ex}{\mbox{\small
{4}}}}\,(v + {\rm H})^{2}\!\left (W^{+}_{\mu\,}W^{-\,\mu} \,+\, \frac{Z_{\mu\,}
Z^{\mu}}{\mbox{\small{2}}\cos^{2\!}\theta_{\mbox{\tiny{W}}}}\right)
\end{equation}
where $D_{\mu}\Phi$ is the {\em covariant derivative} of $\Phi$ and 
\begin{equation}
\cos\theta_{\mbox{\tiny{W}}} = \frac{g}{\sqrt{g^{2} + g'^{\,2}}}~~~~~~~~~~~~~~~
g\sin\theta_{\mbox{\tiny{W}}} = g'\cos\theta_{\mbox{\tiny{W}}} = {\rm e} 
\end{equation}
($g$ and $g'$ are the electroweak couplings and e is the electric charge). A
comparison with the usual mass terms for the charged and neutral vector bosons
reveals that
\begin{eqnarray}
\mbox{\large{$m$}}_{\mbox{\tiny{W}}}^{~}\!\!&=&\!\!\frac{gv}{\raisebox{.2ex}
{\mbox{\small{2}}}} \\*[2mm]
\mbox{\large{$m$}}_{\mbox{\tiny{Z}}}^{~}\!\!&=&\!\!\frac{gv}{\mbox{\small{2}}
\cos\theta_{\mbox{\tiny{W}}}} \,=\, \frac{\mbox{\large{$m$}}_{\mbox{\tiny{W}}}
^{~}}{\cos\theta_{\mbox{\tiny{W}}}}
\end{eqnarray}
From Eq. (14) we also infer that the {\em Higgs-gauge boson couplings} are
\begin{equation}
\lambda_{\mbox{\tiny{HWW}}} \,\equiv\, \frac{g^{2}v}{\raisebox{.2ex}{\mbox
{\small{2}}}} \,=\, \frac{\mbox{\small{2}}\mbox{\large{$m$}}_{\mbox{\tiny{W}}}
^{\,2}}{\raisebox{.5ex}{$v$}}
\end{equation}
and
\begin{equation}
\lambda_{\mbox{\tiny{HZZ}}} \,\equiv\, \frac{g^{2}v}{\mbox{\small{4}}\cos^{2\!}
\theta_{\mbox{\tiny{W}}}} \,=\, \frac{\mbox{\large{$m$}}_{\mbox{\tiny{Z}}}
^{\,2}}{\raisebox{.5ex}{$v$}}
\end{equation}

Therefore, the Higgs couplings to gauge bosons are proportional to their masses.
This can be readily verified by measuring the production cross-sections in the
Higgs-strahlung and W-fusion processes. At center-of-mass energies 
$s \gg \mbox{\large{$m$}}_{\mbox{\tiny{H}}}^{2}$,
\begin{equation}
   \begin{array}{l}
\mbox{\large{$\sigma$}}(e^{+}e^{-} \rightarrow {\rm HZ}) \,\propto\, \lambda
_{\mbox{\tiny{HZZ}}}^{2}/s \\*[4mm]
\mbox{\large{$\sigma$}}(e^{+}e^{-} \rightarrow {\rm H}\nu\bar{\nu}) \,\propto\,
\lambda_{\mbox{\tiny{HWW}}}^{2}\log\!\mbox{\Large{$($}}s/\mbox{\large{$m$}}_
{\mbox{\tiny{H}}}^{\,2}\mbox{\Large{$)$}}
   \end{array}
\end{equation}
The cross-section $\mbox{\large{$\sigma$}}(e^{+}e^{-} \rightarrow {\rm HZ})
\rightarrow {\rm H}\ell^{+}\ell^{-})$ can be measured independently of the
Higgs-boson decay modes by analyzing the invariant mass of the system
recoiling against the Z boson (see Section 5).

The vector bosons are coupled to the ground-state Higgs field by means of the
covariant derivative (see Eq. (14)). The {\em Higgs-fermion couplings} are
introduced in an {\em ad hoc} way through the {\em Yukawa Lagrangian}
\begin{equation}
{\cal L} \,=\, -g_{\mbox{\tiny{$f$}}}^{~}\overline{\psi}_{\mbox{\tiny{$f$}}}^{~}
\psi_{\mbox{\tiny{$f$}}}^{~}\Phi
\end{equation}
Replacing the Higgs field by its ground-state value, $\Phi \rightarrow v/\sqrt
{2}$ (see Eq. (2)), yields the mass term $-\mbox{\large{$m$}}_{\mbox{\tiny
{$f$}}}^{~}\overline{\psi}_{\mbox{\tiny{$f$}}}^{~}\psi_{\mbox{\tiny{$f$}}}
^{~}$, where $\mbox{\large{$m$}}_{\mbox{\tiny{$f$}}}^{~} = g_{\mbox{\tiny
{$f$}}}^{~}v/\sqrt{2}$. The interaction term in the Lagrangian is obtained by
the replacement $\Phi \rightarrow {\rm H}/\sqrt{2}$:
\begin{equation}
{\cal L}_{\rm int} \,=\, -\,\frac{m_{\mbox{\tiny{$f$}}}}{\raisebox{.5ex}{$v$}}
\,{\rm H}\,\overline{\psi}_{\mbox{\tiny{$f$}}}^{~}\psi_{\mbox{\tiny{$f$}}}^{~}
\end{equation}
We see that, in the Standard Model, all the quarks and charged leptons receive
their masses through {\em Yukawa interactions} with the Higgs field.
Note also that the coupling strength between the Higgs field and the fermion 
\mbox{\small{$f$}} is proportional to the mass of the particle.

Using expression (16), as well as
\begin{equation}
\mbox{\large{$m$}}_{\mbox{\tiny{$f$}}}^{~} = \frac{g_{\mbox{\tiny{$f$}}}^{~}v}
{\sqrt{2}}~~~~~~~~~~{\rm and}~~~~~~~~~~\mbox{\large{$m$}}_{\mbox{\tiny{H}}}^{~}
 = \sqrt{2\lambda}~v
\end{equation}
(see Eq. (4)), we obtain\,\footnote{~We can relate $v$ to the Fermi constant
${\rm G}_{\mbox{\tiny{F}}} = 1.16639\times 10^{-5}~{\rm GeV}^{-2}$ as follows:
\[
\frac{{\rm G}_{\mbox{\tiny{F}}}}{\sqrt{\mbox{\small{2}}}} = \frac{g^{2}}{\mbox
{\small{8}}\mbox{\large{$m$}}_{\mbox{\tiny{W}}}^{\,2}} = \frac{\raisebox{-.3ex}
{\mbox{\small{1}}}}{\mbox{\small{2}}v^{2}}~~~\Rightarrow~~~v =\mbox{\large
{$($}}\sqrt{2}\,{\rm G}_{\mbox{\tiny{F}}}\mbox{\large{$)$}}^{-1/2} \approx
246~{\rm GeV} \] }
\begin{equation}
v \,=\, \frac{\mbox{\large{$m$}}_{\mbox{\tiny{W}}}^{~}}{\raisebox{.4ex}{$g$}/
\mbox{\small{2}}}
 \,=\, \frac{\mbox{\large{$m$}}_{\mbox{\tiny{H}}}^{~}}{\sqrt{2\lambda}} 
 \,=\, \frac{\mbox{\large{$m$}}_{\mbox{\tiny{$f$}}}^{~}}{\raisebox{.4ex} 
{$g_{\mbox{\tiny{$f$}}}^{~}$}/\sqrt{\mbox{\small{2}}}}
\end{equation}
This result is illustrated in Fig.\,\ref{fig:Higgs_couplings} for $\mbox{\large
{$m$}}_{\mbox{\tiny{H}}}^{~} = 120$ GeV.

\begin{figure}[t]
\begin{center}
\epsfig{file=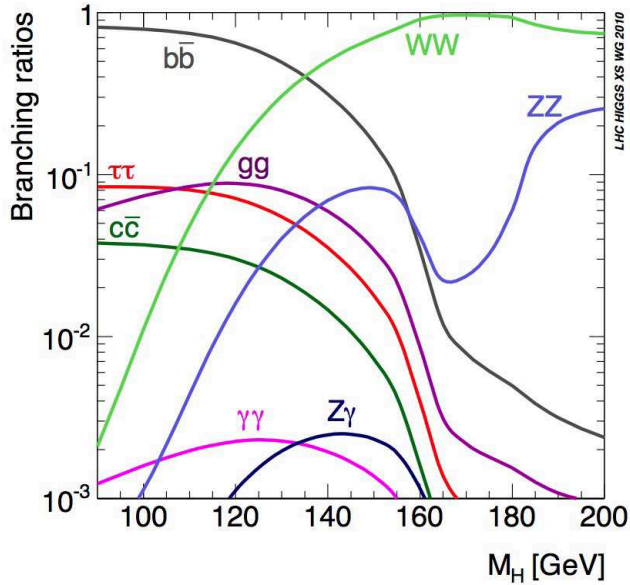,height=0.32\textheight}
\end{center}
\vskip -6mm
\caption{Decay branching fractions of the SM Higgs boson as a function of its
mass. Credit: LHC Higgs XS Working Group.}
\label{fig:BR}
\end{figure}

The Higgs-fermion couplings can be extracted by measuring the {\em branching 
fractions} of the Higgs boson (see Fig.\,\ref{fig:BR}). There are two methods
to determine the Higgs branching fractions: (1) Measure the event rate in the
Higgs-strahlung process for a given final-state configuration and then divide
by the total cross-section; (2) Select a sample of unbiased events in the
Higgs-strahlung recoil-mass peak and determine the fraction of events that
correspond to a particular decay channel. See \cite{heinemeyer} and references
therein for an estimate of the accuracy that can be achieved in such 
measurements. 

For $\mbox{\large{$m$}}_{\mbox{\tiny{H}}}^{~} \gsim 2\mbox{\large{$m$}}_{\mbox
{\tiny{W}}}^{~}$, the {\em total decay width} of the Higgs boson, $\Gamma
_{\mbox{\tiny{H}}}$, is large enough to be determined directly from the
reconstructed Higgs-boson mass spectrum. The result of such an analysis is
shown in \cite{heinemeyer}. For smaller Higgs-boson masses, $\Gamma_{\mbox
{\tiny{H}}}$ can be determined indirectly by employing the relation between
the total and partial decay widths for a given final state:
\begin{equation}
\Gamma_{\mbox{\tiny{H}}} \,=\, \frac{\Gamma ({\rm H} \rightarrow X)}
{{\rm BR}({\rm H} \rightarrow X)}
\end{equation}
For instance, consider the decay ${\rm H} \rightarrow {\rm WW}^{*}$. One can
directly measure the branching fraction ${\rm BR}({\rm H} \rightarrow
{\rm WW}^{*})$, determine the coupling HZZ in the process $e^{+}e^{-}
\rightarrow {\rm HZ}$, relate the HZZ and HWW couplings based on Eqs. 
(18)--(19), and then use the fact that $\Gamma ({\rm H} \rightarrow {\rm WW})
\propto \lambda_{\mbox{\tiny{HWW}}}^{2}$ to obtain the partial width $\Gamma
({\rm H} \rightarrow {\rm WW}^{*})$ from the information on the HWW coupling.
An accuracy between 4\% and 15\% can be achieved in the determination of
$\Gamma_{\mbox{\tiny{H}}}$ for $\mbox{\large{$m$}}_{\mbox{\tiny{H}}}^{~}$ up to
160 GeV \cite{heinemeyer}.

The decay modes ${\rm H} \rightarrow \bar{b}b$,\,WW can also be measured in 
photon-photon collisions with a precision similar to that expected  from 
analyses based on $e^{+}e^{-}$ data (see, e.g., \cite{asner}). Recall from 
Section 4 that the most accurate way to determine the {\em two-photon width}
$\Gamma ({\rm H} \rightarrow \gamma\gamma )$, which is sensitive to the  
Higgs-top coupling, is to combine data from $\gamma\gamma$ and $e^{+}e^{-}$
collisions.

\vspace*{0.3cm}
\section{~The proposed facility in brief}
\vspace*{0.3cm}

~~~~The rich set of final states in $e^{+}e^{-}$ and $\gamma\gamma$ collisions
at a future linear collider (LC) would play an essential role in measuring the 
mass, spin, parity, two-photon width and trilinear self-coupling of the Higgs
boson, as well as its couplings to fermions and gauge bosons; these quantities
are difficult to determine with only one initial state. Furthermore, all the 
measurements made at LEP and SLC could be repeated using highly polarized 
electron beams and at much higher luminosities. For some processes within and 
beyond the Standard Model (e.g., the single and double Higgs-boson production),
the required center-of-mass (CM) energy is considerably lower at the facility
described here than at an $e^{+}e^{-}$ or proton collider. 

A schematic layout of an X-band $e^{+}e^{-}$ linear collider is shown in 
Fig.\,\ref{fig:LC}. Damped and bunch-compressed electron and positron beams 
are accelerated by a pair of linear accelerators (linacs) before colliding at
an interaction point surrounded by a detector. The beams are then disposed of,
and this machine cycle is repeated at a rate of 50 Hz. For a photon collider,
$\gamma\gamma$ collisions are created by Compton backscattering of FEL photons
on high-energy electrons.
 
It is also envisaged that `bypass lines' for low-energy beams would be
employed to accumulate data at the Z resonance in the process $e^{+}e^{-}
\rightarrow{\rm Z}$. These runs could be used to regularly calibrate the
detector, fine-tune the accelerator and measure its luminosity. Assuming a
geometric luminosity $L_{e^{+}e^{-}}^{~} \approx 5\times 10^{33}$ 
cm$^{-2}$\,s$^{-1}$ at the Z resonance, approximately $2\times 10^{9}$ Z bosons 
could be produced in an operational year of $10^{7}$ s; this is about 200 times
the entire LEP statistics.

The proposed facility would be constructed in several stages, each with
distinct physics objectives that require particular center-of-mass (CM)
energies. The processes to be studied, and the corresponding CM energies, are
\cite{belusev2}:
\[
       \begin{array}{ll}
\bullet~~e^{+}e^{-} \rightarrow {\rm Z,\,WW};\hspace*{0.5cm}\gamma\gamma 
\rightarrow {\rm H}~~~~~&~~~~~{\rm E}_{ee} \sim 90~{\rm to}~170~{\rm GeV} 
\\*[4mm]
\bullet~~e^{+}e^{-} \rightarrow {\rm ZH}~~~~~&~~~~~{\rm E}_{ee} \sim 250~
{\rm GeV} \\*[4mm]
\bullet~~\gamma\gamma \rightarrow {\rm HH};\hspace*{0.5cm}e^{+}e^{-} 
\rightarrow t\bar{t}~~~~~&~~~~~{\rm E}_{ee} \sim 330~{\rm to}~350~{\rm GeV}
       \end{array} \]
The top-quark mass and the Higgs-top coupling could be measured in the process
$e^{+}e^{-}\rightarrow t\bar{t}$ at the pair-production threshold \cite{fujii};
one expects $\delta\mbox{\large{$m$}}_{t}^{~} \approx 100~{\rm MeV} \approx
0.1\delta\mbox{\large{$m$}}_{t}^{~}({\rm LHC})$.

The production and testing of the accelerating structures and rf sources needed
for an energy upgrade, and the subsequent installation of the rf sources, can be
carried out with minimal disruption to the data-taking process if the
klystrons, modulators and pulse compressors are placed in a separate tunnel
(see Fig.\,\ref{fig:tunnels}).

There are several notable differences between this and other designs based on
the X-band technology (NLC/JLC, GLC \cite{GLC} and CLICHE \cite{asner}): 
(1) The proposed facility would utilize high-gradient CLIC-type cavities and a 
klystron-based power source; a two-beam scheme could be implemented at a later
stage. (2) There would be only one interaction region and a single beam dumping
system for both $e^{+}e^{-}$ and $\gamma\gamma$ beams. (3) In its first stage of
operation (E$_{ee} \sim 170$ GeV), the entire facility could be placed
{\em within} a site only 3 km long. A facility with these characteristics was
originally proposed in \cite{belusev2} (see also \cite{ruth}).

\vspace*{0.3cm}
\section{~The X-band accelerator complex}
\vspace*{0.3cm}

\begin{figure}[h]
\begin{center}
\epsfig{file=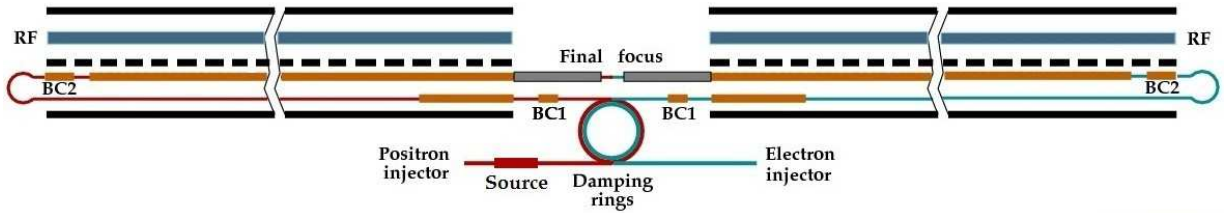,width=0.99\textwidth}
\end{center}
\vskip -7mm
\caption{Schematic layout of an X-band linear $e^{+}e^{-}$ collider.
With a crossing angle at the interaction point (IP), separate beam lines can be
used to bring the disrupted beams to their respective dumps, thereby enabling
post-IP diagnostics. A two-stage bunch compression system (BC) is envisaged.}
\label{fig:LC}
\end{figure}

~~~~A schematic layout of an X-band linear $e^{+}e^{-}$ collider is shown in 
Fig.\,\ref{fig:LC}. The 11.4 GHz X-band rf technology was originally developed
at SLAC and KEK \cite{adolphsen}. The choice of this technology is motivated by
the cost benefits of having relatively low rf energy per pulse and high 
accelerating gradients. The ongoing effort to develop high-gradient X-band 
structures is essential for the eventual construction of a CLIC-type linear
accelerator \cite{ellis}.
 
\begin{figure}[t]
\begin{center}
\epsfig{file=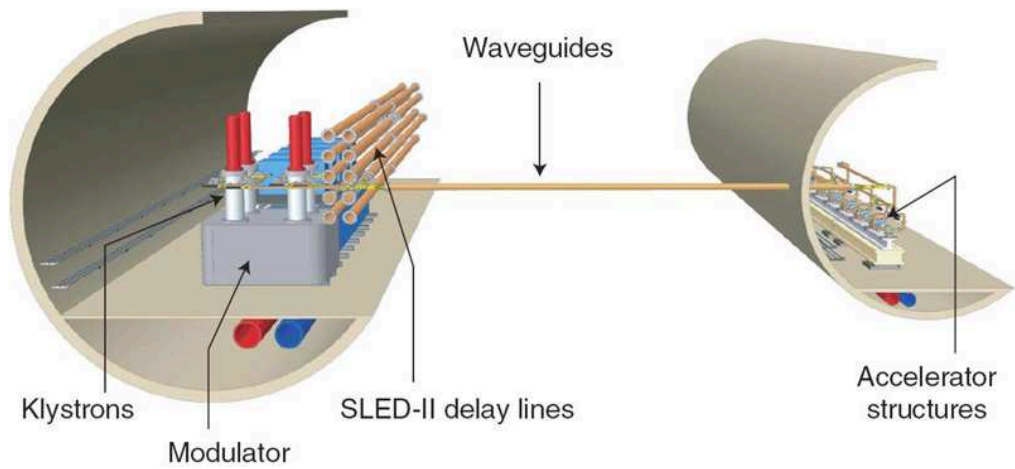,height=0.25\textheight}
\end{center}
\vskip -4mm
\caption{Dual tunnels for an X-band linear collider \cite{GLC}.}
\label{fig:tunnels}
\end{figure}

\begin{figure}[h!]
\begin{center}
\epsfig{file=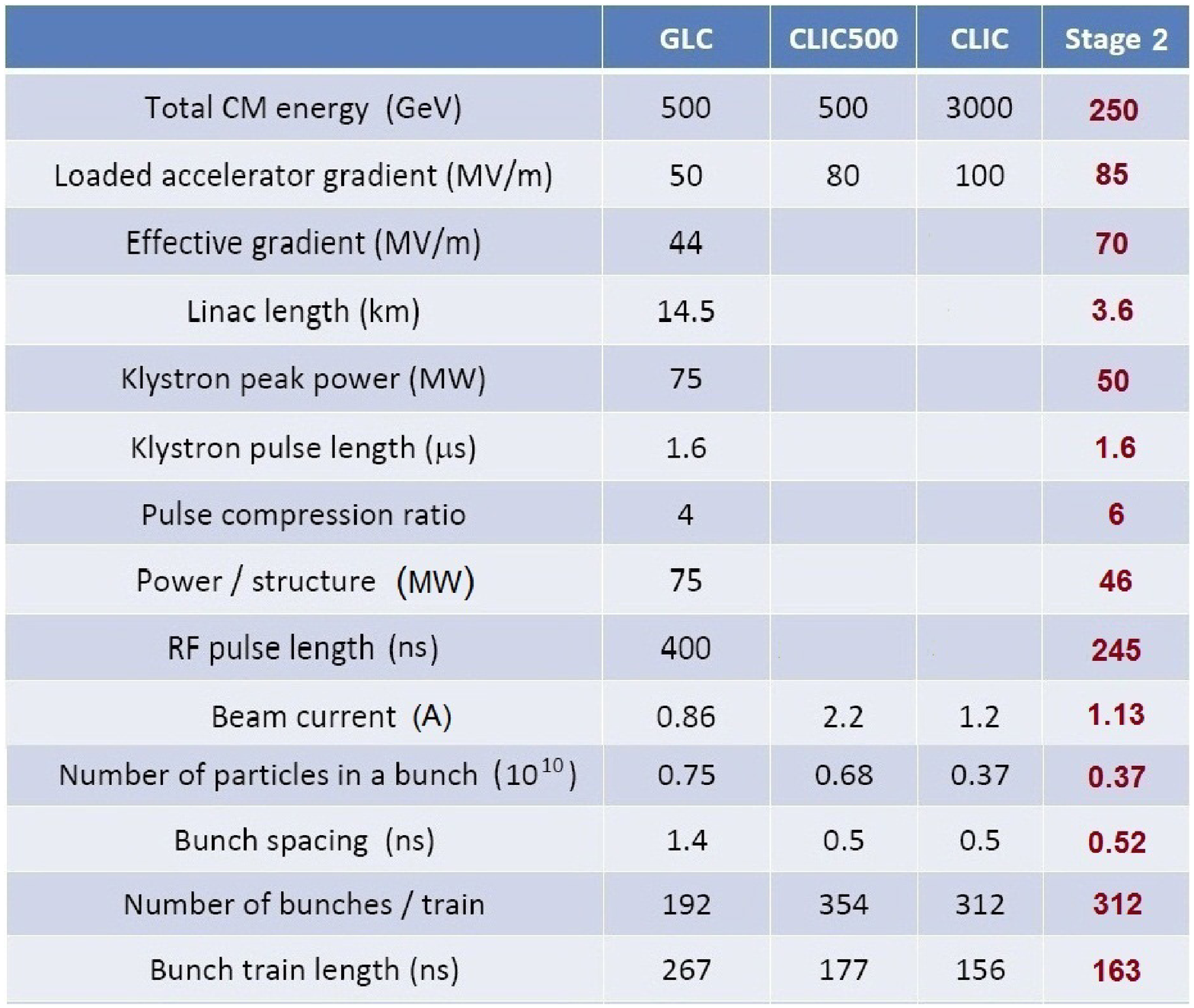,height=0.45\textheight}
\end{center}
\vskip -4mm
\caption{Rough design parameters of the machine.}
\label{fig:Higo1}
\end{figure}

The tunnels containing the rf sources and accelerating structures are sketched
in Fig.\,\ref{fig:tunnels}. As mentioned in Section 8, damped and 
bunch-compressed electron and positron beams are accelerated by a pair of
linacs before colliding at the interaction point. The beams are then disposed
of, and this machine cycle is repeated at a rate of 50 Hz. `Bypass lines' for 
low-energy beams would be employed. Rough design parameters of the machine are
shown in Fig.\,\ref{fig:Higo1} (see \cite{GLC, CLIC}).

A comprehensive review of the status of X-band accelerator technology is given
in \cite{adolphsen}. Since then, significant advances have been made in pulsed 
HV and rf power generation, high gradient acceleration and wakefield supression.
The ultimate design of rf cavities will depend on the outcome of the ongoing
effort to develop 100 MeV/m X-band structures for a CLIC-type linear collider
(see Section 10).

\vspace*{0.3cm}
\section{~The RF system}
\vspace*{0.3cm}

\begin{figure}[h!]
\begin{center}
\epsfig{file=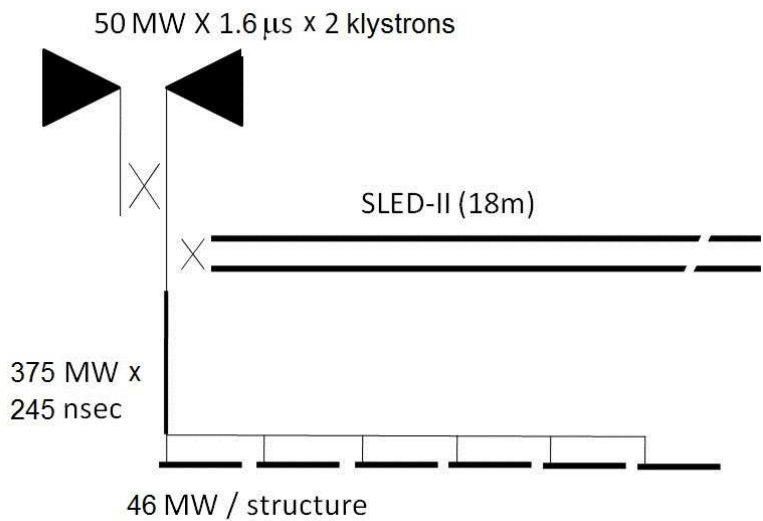,height=0.30\textheight}
\end{center}
\vskip -6mm
\caption{Schematic of the RF acceleration system.}
\label{fig:RF}
\end{figure}

\begin{figure}[h!]
\begin{center}
\epsfig{file=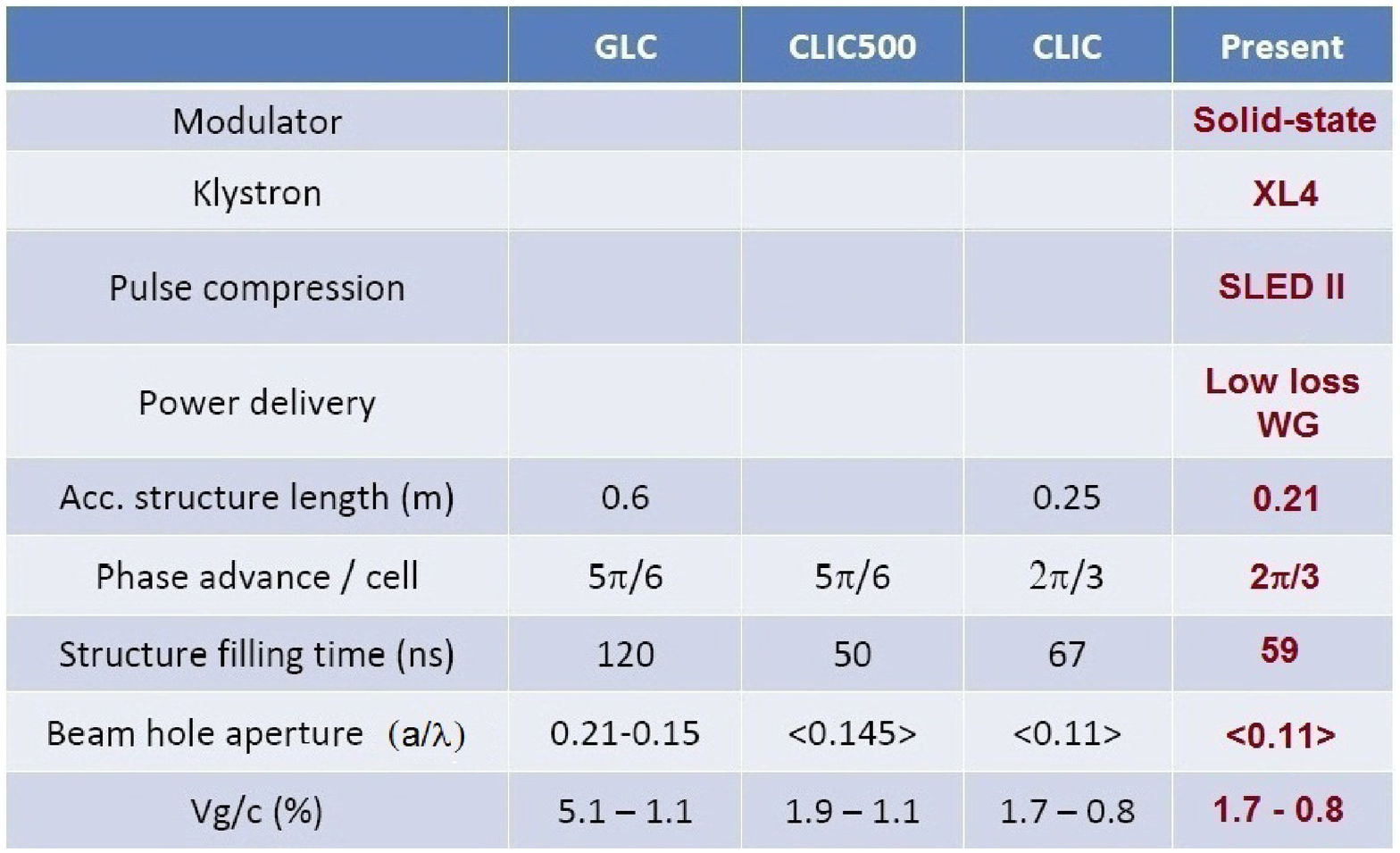,height=0.34\textheight}
\end{center}
\vskip -4mm
\caption{Rough design parameters of the RF system.}
\label{fig:Higo2}
\end{figure}

~~~~A single rf unit contains a modulator that drives a pair of 50 MW 
klystrons, each of which generates 1.6 $\mu$s rf pulses at 50 Hz (see
Figs.\,\ref{fig:RF} and \ref{fig:Higo2}). An rf compression system enhances the
peak power of the klystrons by a factor of 3.75, and produces 245 ns pulses 
that match the accelerator structure requirements.\footnote{~It takes 59.4 ns 
(filling time) plus 22.4 ns (ramping time) to fill each rf cavity with an
accelerating field. The remaining period of 163 ns is used to accelerate a 
`train' of electron bunches.} The resulting 375 MW, 245 ns pulses feed seven
0.21m-long accelerator structures, producing a 85 (100) MV/m loaded (unloaded) 
gradient in each structure. 

\begin{figure}[t]
\begin{center}
\epsfig{file=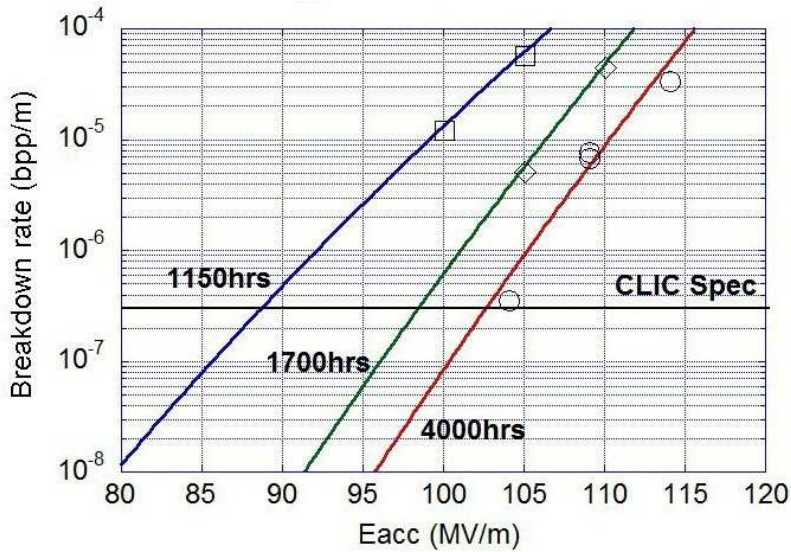,height=0.30\textheight}
\end{center}
\vskip -4mm
\caption{Breakdown rate of the CLIC prototype rf cavity TD24\#4 \cite{higo}.}
\label{fig:Higo3}
\end{figure}

The latest test results for one of the CLIC prototype rf cavities TD24 are 
presented in Fig.\,\ref{fig:Higo3}. The cavity reached the indicated CLIC 
breakdown rate requirement in less than 2000 hours. Based on these results, we
have chosen a value of 85 MeV/m for the beam-loaded accelerator gradient.

To reduce power consumption, it was proposed \cite{ogitsu} to use klystrons
with superconducting solenoidal focusing. The XL4 klystron developed at SLAC,
for instance, could initially be adapted for this purpose. Alternatively, one
could employ klystrons with periodic permanent magnet (PPM) focusing
\cite{sprehn}. However, all PPM klystrons built so far suffer chronic rf
breakdown in the output section, which manifests itself by a loss of
transmitted power that develops over several hundred ns \cite{adolphsen}.

For cost reasons, it is preferable to power as many klystrons per pulse
modulator as possible. With this in mind, a solid-state induction-type
modulator that could drive a pair of 50 MW X-band klystrons was designed at
SLAC \cite{adolphsen}. Another possible choice are solid-state modulators 
produced by ScandiNova. 

\vspace*{0.3cm}
\section{~Photon collider}
\vspace*{0.3cm}

~~~~The idea of using counter-directed electron linacs to create a gamma-gamma
collider can be traced back to an article by P. Csonka\footnote{One of the
present authors (R. Belusevic) has maintained a keen interest in Paul Csonka's
pioneering idea since the late 1970s.} published in 1967 \cite{csonka}. The 
seminal work on photon colliders by I. Ginzburg et al. \cite{ginzburg} describes
in detail a method for obtaining $\gamma\gamma$ and $e\gamma$ collisions by 
Compton backscattering of laser light on high-energy electrons. 

\begin{figure}[t]
\begin{center}
\epsfig{file=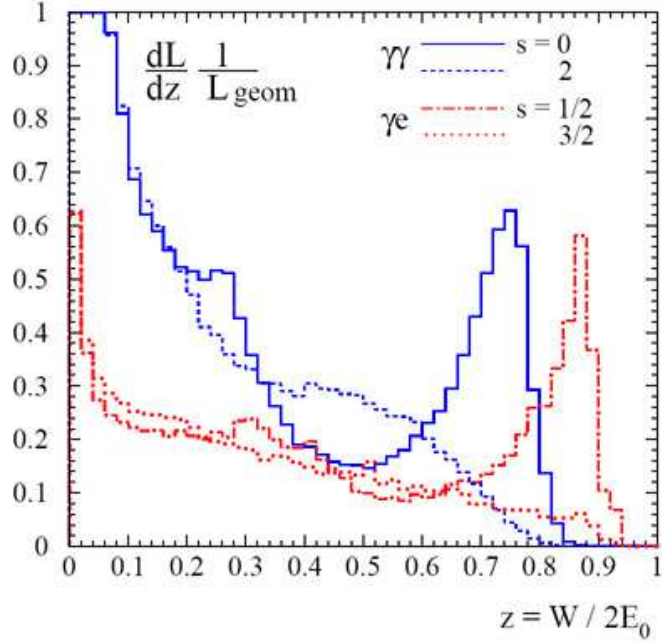,height=0.35\textheight}
\end{center}
\vskip -7mm
\caption{Simulated $\gamma\gamma$ and $e\gamma$ luminosity spectra
\cite{telnovACTA}.}
\label{fig:spectra}
\end{figure}

The backscattered photons have energies comparable to those of the incident
electrons (see Fig.\,\ref{fig:spectra}), and follow their direction with some
small angular spread of the order of $1/\gamma$, where $\gamma$
is the Lorentz factor. The spatial spread of the photons is
approximately $d/\gamma$ at a distance $d$ from the Compton interaction point
(CIP). Both the energy spectrum and polarization of the backscattered photons
depend strongly on the polarizations of the incident electrons and laser
photons. The key advantage of using $e^{-}e^{-}$ beams at a $\gamma\gamma$
collider is that they can be polarized to a high degree.

At CIP, the electron beam is about 10 times wider than it would be at the $ee$
collision point in the absence of a laser beam. However, since the
backscattered photons follow the direction of the incident electrons, they are
automatically `focused' to their collision point.

The absence of beam-beam effects in $\gamma\gamma$ collisions means that it
is not necessary to have very flat linac beams. The spectral luminosity of
$\gamma\gamma$ collisions strongly depends on beam characteristics, but only
through the parameter $\rho$, the ratio of the intrinsic transverse spread of
the photon beam to that of the original electron beam: $\rho \equiv d/\gamma
\sigma_{e}^{~}$. In this expression, $d$ is the distance between CIP and the
photon-photon collision point, $\gamma$ is the Lorentz factor and  
$\sigma_{e}^{~}$ is the radius that a round Gaussian linac beam would have at
the collision point in the absence of a laser beam. As $\rho$ increases, the
the monochromaticity of the luminosity distribution improves (because the
lowest-energy photons, which scatter at the largest angles, do not pass through
the collision point), but the total luminosity decreases. For a typical photon 
collider, the optimal value of $d$ is a few millimeters \cite{TESLA}. 

Assuming that the mean number of Compton interactions of an electron in a laser
pulse (the Compton conversion probability) is 1, the {\em conversion
coefficient} $k \equiv n_{\gamma}^{~}/n \approx 1 - \mbox{\small{e}}^{-1} =
0.63$, where $n_{e}$ is the number of electrons in a 'bunch' and $n_{\gamma}
^{~}$ is the number of scattered photons. The luminosity of a gamma-gamma
collider is then
\begin{equation}
{\cal L}_{\gamma\gamma} \,=\, (n_{\gamma}^{~}/n_{e}^{~})^{2\,}{\cal L}_{ee}
 \,\approx\, (0.63)^{2\,}{\cal L}_{ee}
\end{equation}
where ${\cal L}_{ee}$ is the {\em geometric luminosity} of electron
beams:
\begin{equation}
{\cal L}_{ee} \,=\,\frac{\gamma n_{e}^{\,2}{\rm N}_{b}f}{4\pi\sqrt{\mbox{\large
{$\varepsilon$}}_{x}\beta_{x}\mbox{\large{$\varepsilon$}}_{y}\beta_{y}}}
\end{equation}
In this expression, $\mbox{\large{$\varepsilon$}}_{x},\mbox{\large
{$\varepsilon$}}_{y}$ are the {\em beam emittances}, $\beta_{x},\beta_{y}$ are
the horizontal and vertical {\em beta functions}, respectively, N$_{b}$ is the
number of bunches per train, and $f$ is the beam collision frequency. In
the high-energy part of the photon spectrum, ${\cal L}_{\gamma\gamma}\sim 0.1
{\cal L}_{ee}$. However, if beams with smallest possible emittances and
stronger beam focusing in the horizontal plane are used, then ${\cal L}_{\gamma
\gamma}$ could, in principle, exceed ${\cal L}_{e^{+}e^{-}}$ \cite{telnovACTA}
(see also Section 14).

\vspace*{0.3cm}
\section{~Laser system for an X-band machine}
\vspace*{0.3cm}
 
~~~~In order to attain maximum luminosity, every electron bunch in the 
accelerator should collide with a laser pulse of sufficient intensity for
$63\%$ of the electrons to undergo a Compton scattering. This requires a laser
system with high average power, capable of producing pulses that would match
the temporal spacing of electron bunches. The laser power is minimized when 
the Rayleigh range of the laser focus and the laser pulse width are matched to
the electron bunch length. The proposed collider would have about 300 50-micron
bunches separated by 0.5 ns, with 50 trains per second. This means that 
$300\times 50 = 15000$ laser pulses with a duration of approximately 1 ps must
be produced every second. To avoid nonlinear electrodynamic effects, the
maximum pulse energy should not exceed a couple of joule. Therefore, the laser 
system ought to deliver at least 15 kW of average power in pulses of a terawatt 
peak power, matched to the linac bunch structure. 

These requirements could be satisfied by an optical {\em free electron laser} 
(FEL). The radiation produced by an FEL has a variable wavelength, and is fully
polarized either circularly or linearly depending on whether the undulator is
helical or planar, respectively. The required time structure of laser pulses
can be achieved by using an S-band linac for the FEL. A free electron laser for
a photon collider is described in \cite{saldin}.

The wavelength $\lambda$ of FEL radiation is determined by $\lambda \approx 
\lambda_{u}/2\gamma^{2}$, where $\gamma$ is the Lorentz factor of the electron
beam and $\lambda_{u}$ is the periodic length of the undulator. An optical FEL
requires a much smaller electron linac and a considerably simpler undulator
than an XFEL. However, the charge per electron beam bunch has to be 
sufficiently large ($\sim 5$ nC) to produce photon pulses of $\sim 1$ J. 
Suitable high-intensity and low-emittance rf guns have already been developed
\cite{michelato}. An optical FEL could be placed in a separate tunnel
connected to the experimental hall housing the detector. 

\vspace*{0.3cm}
\section{~Interaction region and beam dump}
\vspace*{0.3cm}

~~~~The location of beamline elements near the interaction region (IR) of an 
$e^{+}e^{-}$ collider and their integration with a generic detector are 
discussed at length in \cite{CLIC} and \cite{ILC}.

\begin{figure}[!t]
\begin{center}
\epsfig{file=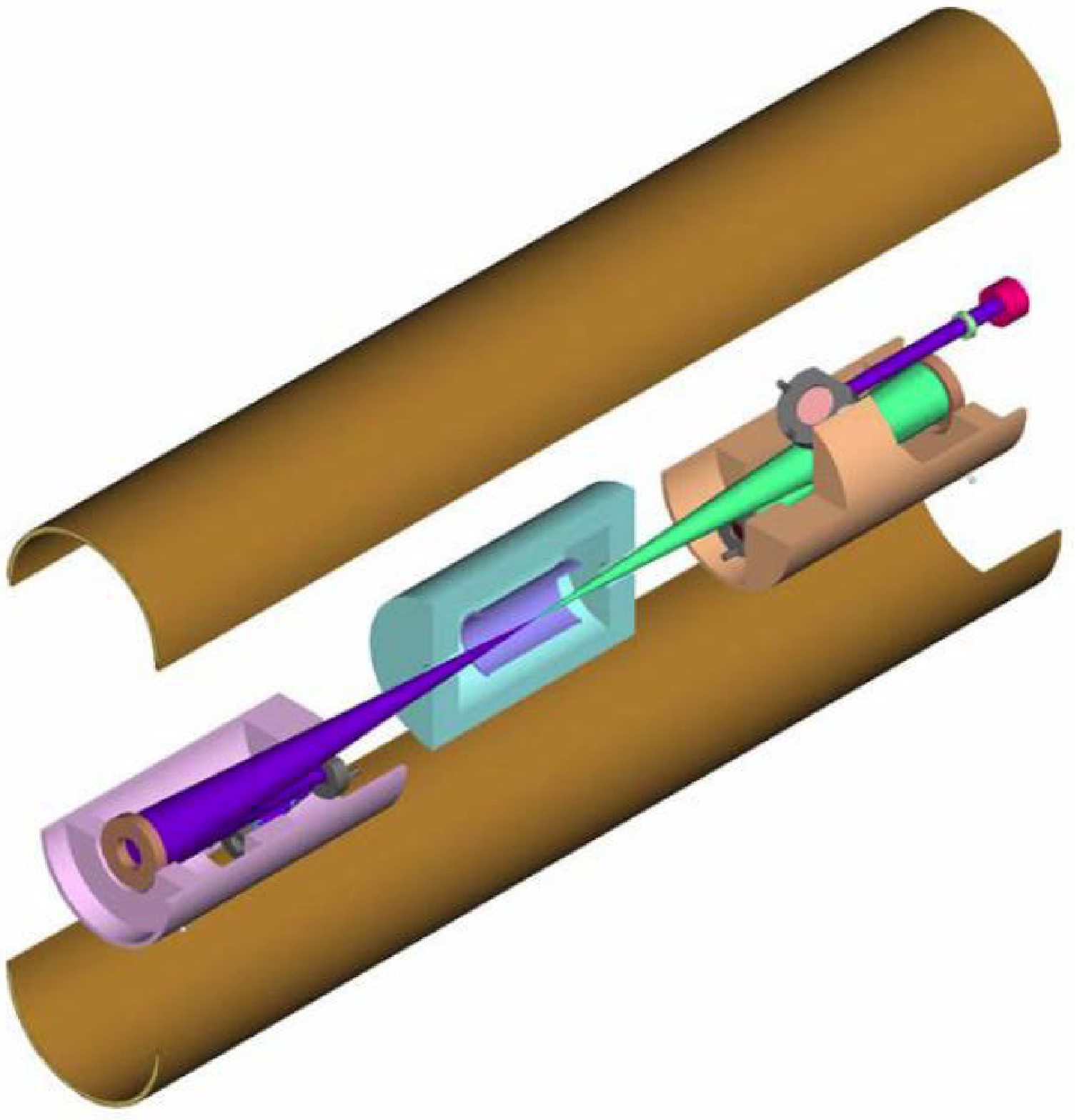,height=0.33\textheight}
\epsfig{file=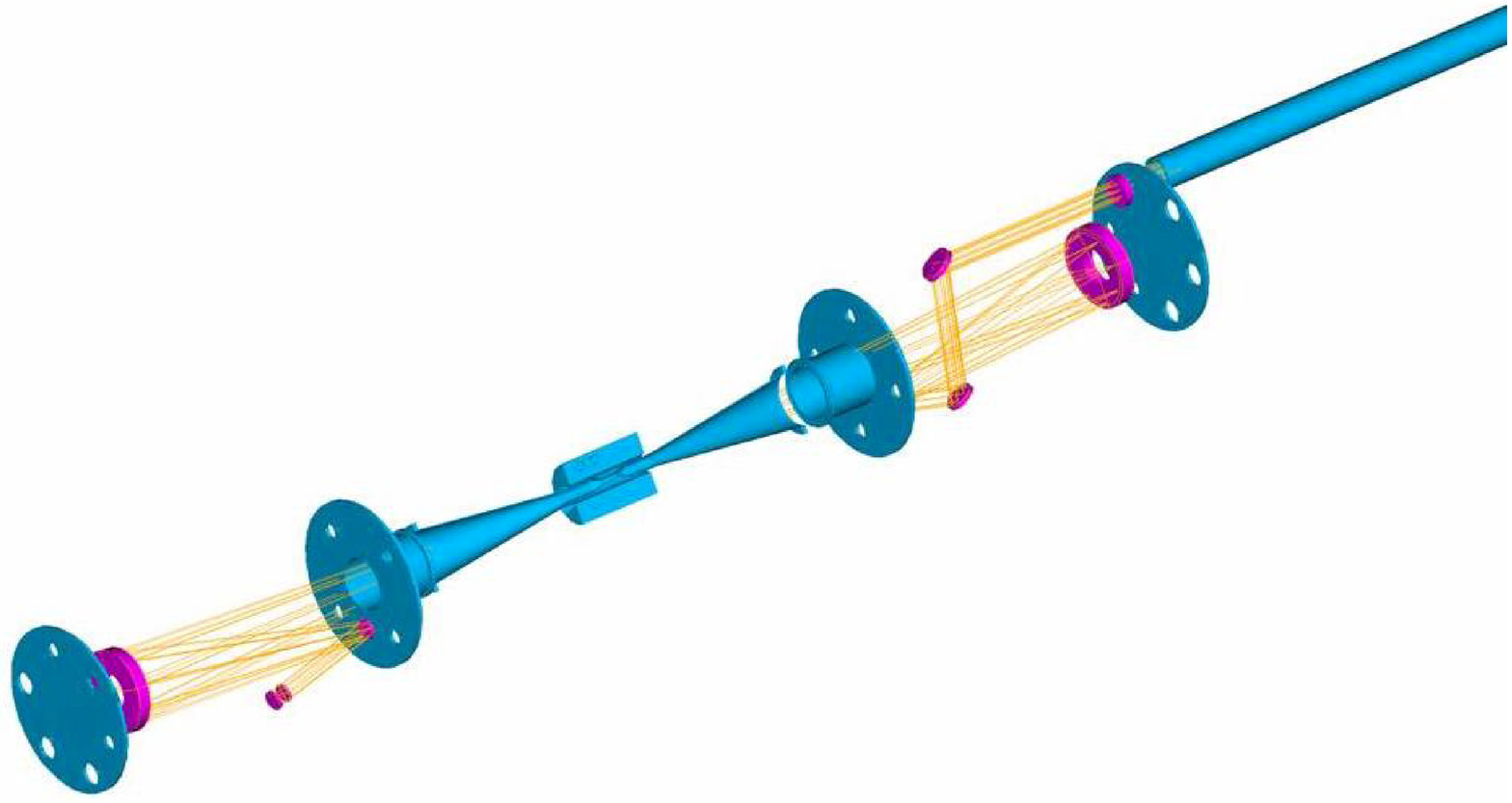,height=0.24\textheight}
\end{center}
\vskip -4mm
\caption{Optics assembly at the $\gamma\gamma$ interaction region. Electron
beams and most of the background particles pass through the central hole in
each of the two end-mirrors. Elements of the assembly were designed, simulated
and prototyped at LLNL \cite{skulina, gronberg1}.}
\label{fig:optics}
\end{figure}

The {\em assembly} for the interaction region at a photon collider shown in 
Fig.\,\ref{fig:optics} satisfies the following requirements:
(a) the laser beam must be nearly co-linear with the electron beam;
(b) the latter must pass through the final focusing optics;
(c) the beams of electrons and laser photons must simultaneously be at the
Compton interaction point;
(d) the duration of the laser pulse must correspond to the electron bunch 
length. 

The Compton scattering of laser photons on high-energy electrons results in a
large energy spread in the electron beam. At the interaction point (IP), this 
leads to a large angular spread of the outgoing beam due to the beam-beam 
interaction. For nominal beam and laser parameters, the extraction beam pipe
must therefore have an aperture of about $\pm 10$ milliradians.

To remove the disrupted beams, one can use the {\em crab-crossing
scheme} proposed by R. Palmer. In this scheme, the beams are collided at a
{\em crossing angle} of about 10 to 20 milliradians. The same luminosity as in
head-on collisions can be obtained by tilting the electron bunches with respect
to the direction of the beam motion. 

The aperture of the extraction beam pipe and the physical size of the final
focusing magnet set a lower limit on the crossing angle of the colliding beams.
The minimum crossing angle is about 25 milliriadians if a final focusing
quadrupole magnet with a compensating coil to minimize the fringe field is used
\cite{parker}.\,\footnote{~The fringe field from the final focusing magnet must
be minimized to prevent low-energy particles, which are swept away by the
field, from causing radiation- and heat-related problems.} This is somewhat 
larger than the crossing angles envisaged for the proposed GLC and NLC X-band 
$e^{+}e^{-}$ colliders.

The `feedback' system for bringing the beams into collision relies on post-IP 
{\em beam position monitors} (BPMs) that measure the beam-beam deflection
at the collision point. Because of the energy spread in a highly disrupted 
beam, conventional BPMs may not provide sufficient resolution due to electric
noise. Moreover, it is not possible to steer such a beam without large beam
losses. This implies that the extraction line at a $\gamma\gamma$ collider will
be a straight vacuum pipe, which precludes some post-IP diagnostics such as
precise measurement of the beam energy and polarization. 


Much of the extracted-beam power will be in the form of high-energy photons 
that have a very narrow angular spread. This would result in a large amount of
energy being deposited within a small volume of the water {\em beam dump}, 
causing vaporization of H$_{2}$O. A possible solution to this problem
would be to convert the photons to $e^{+}e^{-}$ pairs in a gas target situated
before the dump. In order to decrease the flux of backward-scattered neutrons,
a volume filled with hydrogen or helium gas could be placed just before the gas
target (see \cite{shekhtman} and \cite{telnovACTA} for more detail).

Huge savings in construction cost could be achieved if the crossing angle and
the beam dump are exactly the same for the operation of the accelerator in the
$e^{+}e^{-}$ and $\gamma\gamma$ collision modes. The beam dump described in
\cite{shekhtman} is designed with this in mind. The part of the extraction line
containing a chicane --- which provides vertical displacement and dispersion 
needed for continuous measurements of the beam energy spectrum and polarization
at an $e^{+}e^{-}$ collider --- could be replaced with a straight vacuum pipe 
for the operation in the $\gamma\gamma$ mode. 

\vspace*{0.3cm}
\section{~Luminosity and backgrounds at a $\gamma\gamma$ collider}
\vspace*{0.3cm}

~~~~Since the cross-sections 
$\mbox{\large{$\sigma$}}_{\gamma\gamma\,\rightarrow\,\mbox{\tiny{HH}}}$ and
$\mbox{\large{$\sigma$}}_{e^{+}e^{-}\,\rightarrow\,\mbox{\tiny{HHZ}}}$ do not 
exceed 0.4 fb, it is essential to attain the highest possible luminosity,
rather than energy, in order to measure the trilinear Higgs self-coupling. If
beams with smallest possible emittances and stronger beam focusing in the
horizontal plane are used, then the $\gamma\gamma$ luminosity ${\cal L}_{\gamma
\gamma}$ could, in principle, be made higher than ${\cal L}_{e^{+}e^{-}}$, as
explained in what follows \cite{telnovACTA}.

At a photon collider with CM energies $\lsim 500$ GeV, and for electron beams
that are not too short, coherent pair production is suppressed due to the
broadening and displacement of the electron beams during their collision.
In this case, ${\cal L}_{\gamma\gamma}$ is limited {\em only} by the 
transverse area of the beam (note that its vertical size is much smaller
than the horizontal):
\begin{equation}
{\cal L}_{\gamma\gamma} \,\propto\, (\mbox{\large{$\sigma$}}_{x}^{~}\mbox
{\large{$\sigma$}}_{y}^{~})^{-1}~~~~~~~~~~~~~~~\mbox{\large{$\sigma$}}_{x,y}
^{~} \,=\, \sqrt{\beta_{x,y}(\mbox{\large{$\varepsilon$}}_{x,y}/\gamma )}
\end{equation}
as can be seen from expressions (26) and (27) in Section 11.

\begin{figure}[!t]
\begin{center}
\epsfig{file=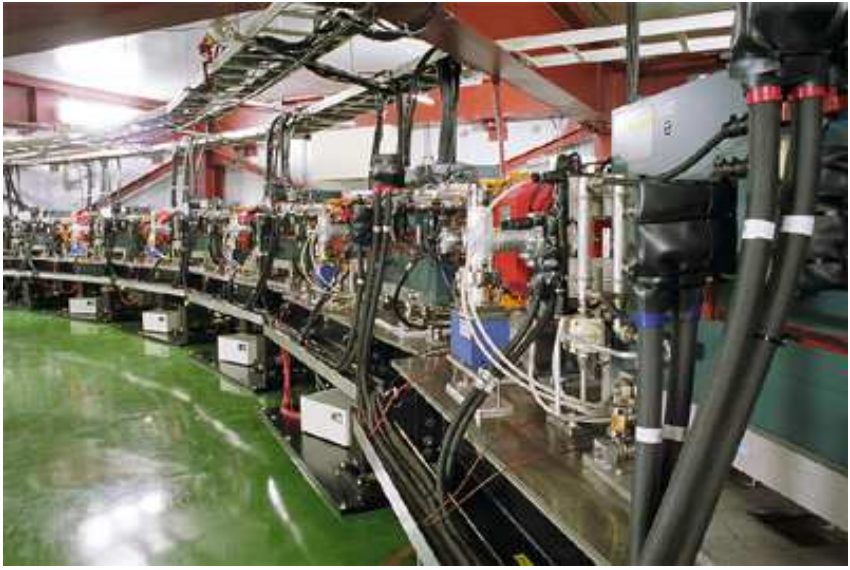,height=0.32\textheight}
\end{center}
\vskip -4mm
\caption{Arc section of the ATF Damping Ring, which produces the world's
smallest-emittance beams. The layout of the magnets in the arc sections is
designed to achieve small equilibrium emittances. The wiggler magnets in the
straight sections of the ring shorten the damping time. Fast kicker magnets and
DC septum magnets are used for beam extraction. Credit: ATF team.}
\label{fig:ATF}
\end{figure}

The beam emittances in Eq. (28) are determined by various physics effects
inside a damping ring (see Fig.\,\ref{fig:ATF}). If the synchrotron radiation
is dominated by the ring's wiggler parameters (large $F_{\rm w}$), and if the
quantum excitation by the wiggler is not too large compared with that in the
arcs, then from Eqs. (33) and (14) in \cite{emma} it follows that the
horizontal beam emittance $\mbox{\large{$\varepsilon$}}_{x}$ could be
significantly reduced by using a wiggler with short period and a judicially
chosen value of the peak field (in order to preserve the damping time). The
vertical emittance $\mbox{\large{$\varepsilon$}}_{y}$ is {\em not} determined
by the wiggler, but by optics errors that are not easily characterized.
Assuming, for instance, that $\mbox{\large{$\varepsilon$}}_{x}$ could be 
reduced by a factor of 6 and $\mbox{\large{$\varepsilon$}}_{y}$ by a factor of
4 compared with their `nominal' ILC values, ${\cal L}_{\gamma\gamma}$ would 
then exceed ${\cal L}_{e^{+}e^{-}}$: 
\[ {\cal L}_{\gamma\gamma}(\mbox{\small{high-energy peak}}) \,\sim\, 
1.2\,{\cal L}_{e^{+}e^{-}} \]
where ${\cal L}_{e^{+}e^{-}} \approx 2\times 10^{34}\,{\rm cm}^{-2}\,{\rm s}
^{-1}$ is limited by collision effects (beamstrahlung and beam instabilities)
\cite{telnovACTA}. To obtain this result it was also assumed that $\beta_{x} = 
1.7$\,mm and that the distance between the interaction and conversion $d = 1$ 
mm. Note that the minimum value of $\beta_{x}$ is restricted by
chromo-geometric aberrations in the final-focus system. Simulated luminosity 
spectra for these values of $\beta_{x}$ and $d$ are shown in 
Fig.\,\ref{fig:spectra}. 

At a $\gamma\gamma$ collider, the spectrum of photons after Compton scattering
is broad, with a characteristic peak at maximum energies (see 
Fig.\,\ref{fig:spectra}). The low-energy part of the spectrum is produced by 
multiple Compton scattering of electrons on photons inside laser beams. 

The Compton-scattered photons can have circular or linear polarizations,
depending on their energies and the polarizations of the initial electrons and
laser light. For instance, the scattered photons have an average helicity
$\langle\lambda_{\gamma}\rangle \neq 0$ if either the laser light has a 
circular polarization ${\rm P}_{\!c} \neq 0$ or the incident electrons have a
mean helicity $\langle\lambda_{e}\rangle \neq 0$. In the case $2{\rm P}_{\!c}
\lambda_{e}= -1$, which results in a good monochromaticity of the backscattered
photon beam, the average degree of circular polarization of the photons within
the high-energy peak of the luminosity distribution is over 90\%.

Since the polarization of Compton-scattered photons depends strongly on their
energy, the {\em luminosity spectrum} has to be measured separately for 
different polarization states. When both photons are {\em circularly 
polarized}, the process $\gamma\gamma\rightarrow e^{+}e^{-},\,\mu^{+}\mu^{-}$ 
is particularly well suited for measuring the spectral luminosity \cite{pak}. 
This process has a cross-section of a few pb for a total $\gamma\gamma$ angular
momentum $|J_{z}| = 2$. A precision of about 0.1\% is expected in one year of
running, which is better than the accuracy needed for the Higgs-boson studies 
described in this note. To measure the luminosity spectrum in the $|J_{z}| = 0$
configuration, the helicity of one of the photon beams can be inverted by
simultaneously changing the signs of the helicities of both the laser and 
electron beams. For the product of photon {\em linear polarizations}, the
spectral luminosity can be measured in the above process by studying the 
azimuthal variation of the cross-section at large angles \cite{pak, TESLA}.

Undisrupted electron beams at a $\gamma\gamma$ collider can be steered using a 
fast feedback system that measures their deflection (see Section 13). Once the 
electron beams are brought into collision, the laser will be turned on. The
scattered photons follow the direction of the incident electrons.

Multiple Compton scattering of electrons on photons leads to a low-energy
`tail' in the energy spectrum of the electrons. At the interaction point, this
results in a large deflection angle of the $e^{-}e^{-}$ beams. Due to a finite
crossing angle (see Section 13), the outgoing beams are also deflected 
vertically by the solenoidal magnetic field of the detector. Fig.\, 19 in
\cite{bechtel} shows the angular spread of an outgoing electron beam right
after the interaction point and at $z = 2.8$ m. The problem of 'stabilizing'
beam-beam collisions, and hence the $\gamma\gamma$ luminosity, is discussed in 
\cite{telnovACTA}. 

The {\em backgrounds} at a photon collider caused by beam-beam effects 
in the interaction region have been simulated considering both the 
{\em incoherent} particle-particle and {\em coherent} particle-beam
electromagnetic interactions \cite{TESLA}. Another
significant source of background is due to backscattering of particles. The
hadronic structure of the photon arises from the possibility that it can either
split into a quark-antiquark pair or transform into a vector meson, with the
probability of about 1/200. At the expected ILC $\gamma\gamma$ luminosity, for
instance, the average number of hadronic background events per one bunch 
collision is about two \cite{TESLA}. The above backgrounds influence data
acquisition and analysis, as well as the operation of various detector
components \cite{TESLA, bechtel}. 

\newpage

\section{~Main advantages over a TESLA-type design}
\vspace*{0.3cm}

~~~~A detailed description of the {\em International Linear Collider} (ILC)
design can be found in \cite{ILC}. This design, based on the
superconducting technology originally developed at DESY, uses L-band (1.3 GHz)
rf cavities that have average accelerating gradients of $\sim 31.5$ MeV/m. A
single superconducting niobium cavity is about 1 m long. Nine cavities are
mounted together in a string and assembled into a common low-temperature
cryostat or {\em cryomodule}, which is 12.652 m long.

The ILC main linacs are composed of rf units, each of which is formed by three
contiguous cryomodules containing 26 nine-cell cavities. Every unit has an rf
source, which includes a pulse modulator, a 10 MW multi-beam klystron, and a
waveguide system that distributes the power to the cavities.

This L-band (TESLA-type) design offers some advantages over the X-band
technology:

\vspace*{2mm}
\noindent
$\bullet$~~Wakefields are considerably reduced due to the large size of the rf
cavities, which means that cavity alignment tolerances\footnote{In the main 
linac of a linear collider, the principal sources of emittance growth are
misaligned quadrupoles (which introduce dispersion) and off-axis accelerating 
cavities (which generate transverse wakefields).} can be relaxed;\\
\noindent
$\bullet$~~Superconducting rf cavities can be loaded using a long rf pulse
(1.5 ms) from a source with relatively low peak rf power;\\
\noindent
$\bullet$~~`Wall-plug to beam' power transfer efficiency is considerably higher 
than for a klystron-based X-band machine;\\  
\noindent
$\bullet$~~The long rf pulse allows a long bunch train ($\sim 1$ ms), with many
bunches ($\sim 2600$) and a relatively large bunch spacing ($\sim 370$ ns). A
trajectory correction (feedback) system within the train can therefore be used
to bring the beams into collision.
\vspace*{2mm}

However, in contrast to a compact, high-gradient X-band machine, a collider
based on the current TESLA-type design would be characterized by low
accelerating gradients, damping rings whose length (at least 6 kilometers)
limits the luminosity of the machine, and a technologically challenging
cryogenic system comprising a number of surface cryogenic plants. Such plants
require access roads and an electric power-supply network, and have to be 
connected to the accelerator via horizontal or vertical shafts. 
 
Low-gradient, TESLA-type linacs are bound to be very long. A considerable
fraction of the total cost of the accelerator would therefore be spent on civil
engineering, whereas for an X-band machine it would be spent mainly on the 
accelerating structures and rf sources. But the most serious drawback, in our
opinion, of the TESLA-type design is that it cannot be used as a prototype for
a high-gradient, TeV-scale linear collider such as CLIC. 

It is also important to bear in mind that the electron bunches in a TESLA-type
machine are considerably longer than in an X-band linac, which means that more
powerful laser pulses are needed to achieve comparable photon densities at the
Compton interaction region.
 
\vspace*{0.3cm}
\section{~Summary and Acknowledgements}
\vspace*{0.3cm}

~~~~The rich set of final states in $e^{+}e^{-}$ and $\gamma\gamma$ collisions
at a future linear collider would play an essential role in measuring the mass,
spin, parity, two-photon width and trilinear self-coupling of the Higgs boson,
as well as its couplings to fermions and gauge bosons (see Sections 4 to 6);
these quantities are more difficult to determine with only one initial state. 
For some processes within and beyond the Standard Model, the required 
center-of-mass energy is considerably lower at the facility described here than
at an $e^{+}e^{-}$ or proton collider.

Since the cross-sections
$\mbox{\large{$\sigma$}}_{\gamma\gamma\,\rightarrow\,\mbox{\tiny{HH}}}$ and
$\mbox{\large{$\sigma$}}_{e^{+}e^{-}\,\rightarrow\,\mbox{\tiny{HHZ}}}$ do not
exceed 0.4 fb, it is essential to attain the highest possible luminosity,
rather than energy, in order to measure the trilinear Higgs self-coupling. If
beams with smallest possible emittances and stronger beam focusing in the
horizontal plane are used, then the luminosity ${\cal L}_{\gamma\gamma}$ could,
in principle, exceed ${\cal L}_{e^{+}e^{-}}$ (see Section 14).

The proposed $e^{+}e^{-}/\gamma\gamma$ collider would be constructed in several
stages, each with a distinct physics objective that requires a particular 
center-of-mass energy (see Section 8 and the preprint in \cite{belusev1}).
Together with LHC, such a facility would bridge the gap between the present 
high-energy frontier and that accessible to a TeV-scale $e^{+}e^{-}$ or muon
collider. Moreover, the proposed facility would also serve as a prototype for a 
linear collider with high accelerating gradients based on the CLIC design.

If the initial operation of the proposed facility is in the $\gamma\gamma$
mode, there would be no need for an $e^{+}$ source. Two electron damping
rings could then be built inside a single tunnel. For operation at the nominal 
$e^{+}e^{-}$ luminosity, a positron damping ring would later replace one of the
electron rings.

A possible source of primary photons for a $\gamma\gamma$ collider is an  
optical {\em free electron laser} (FEL). The radiation produced by an FEL 
has a variable wavelength and is fully polarized either circularly or 
linearly. Each of the three currently operating X-ray free electron lasers 
(XFELs) --- at SLAC (S-band), DESY (L-band) and the SPring-8 facility (C-band)
--- can serve as a testbed for an optical FEL.

Elements of the optics assembly for the interaction region at a photon collider
were designed, simulated and prototyped at LLNL (see Section 13).
The Compton scattering of laser photons on high-energy electrons results in a
large energy spread in the electron beam. At the interaction point, this
leads to a large angular spread of the outgoing beam due to the beam-beam
interaction. To remove the disrupted beams, one can use the crab-crossing 
scheme described in Section 13. Huge savings in construction cost could be
achieved if the crossing angle and the beam dump are exactly the same for the
operation of the accelerator in the $e^{+}e^{-}$ and $\gamma\gamma$ collision 
modes.

An L-band (TESLA-type) linear collider offers some advantages over an X-band
machine (see Section 15). However, in contrast to a compact, high-gradient
X-band accelerator, a collider based on the current TESLA-type design would be
characterized by low accelerating gradients, damping rings whose length (a
few kilometers in circumference) limits the luminosity of the machine, and a
technologically challenging cryogenic system that requires a number of surface
cryogenic plants. Furthermore, the electron bunches in a TESLA-type machine are
considerably longer than in an X-band linac, which means that much more  
powerful laser pulses are needed to achieve comparable photon densities at the
Compton interaction region. But the most serious drawback, in our opinion, of
the TESLA-type design is that it cannot be used as a prototype for a 
high-gradient, TeV-scale linear collider such as CLIC.

\vspace*{0.8cm}
\begin{center}
{\large\bf Acknowledgements}
\end{center}
\vspace*{0.4cm}

The study of high-gradient rf structures mentioned in Section 10 of this
proposal was initiated by the CLIC collaboration, and has been conducted at
CERN, KEK and SLAC. We greatly appreciate the support we have received from
W. Wuensch and his colleagues at CERN, and from S. Tantawi and his 
collaborators at SLAC.

We are also grateful to K. Fl\"{o}ttmann, K. Fujii, S. Fukuda, S. Hiramatsu, 
G. Jikia, S. Kazakov, K. Kubo, S. Matsumoto, A. Miyamoto, K. Oide, Y. Okada, 
A. Seryi, D. Sprehn, N. Toge, A. Wolski and K. Yokoya for their help with 
regard to various aspects of this proposal.


\newpage
 
\begin{thebibliography}{99}
\vspace*{0.5cm}


\bibitem{GLC} 
GLC Project Report, K. Abe et al. (2003).

\bibitem{LEP} 
The LEP Collaborations, R. Barate et al., Phys. Lett. B 
{\bf 565}, 61 (2003). 

\bibitem{erler}
J. Erler et al., Phys. Lett. B {\bf 486}, 125 (2000).


\bibitem{h0-mass1}
H.~E.~Haber and R.~Hempfling,
Phys.\ Rev.\ Lett.\  {\bf 66}, 1815 (1991);
Y.~Okada, M.~Yamaguchi and T.~Yanagida,
Prog.\ Theor.\ Phys.\  {\bf 85}, 1 (1991);
J.~R.~Ellis, G.~Ridolfi and F.~Zwirner,
Phys.\ Lett.\ B {\bf 257}, 83 (1991).

\bibitem{h0-mass2}
M.~Carena et al., Phys.\ Lett.\ B {\bf 355}, 209 (1995);
M.~Carena, M.~Quiros and C.~Wagner,
Nucl.\ Phys.\ B {\bf 461}, 407 (1996);
H.~E.~Haber, R.~Hempfling and A.~Hoang,
Z.\ Phys.\ C {\bf 75}, 539 (1997);
S.~Heinemeyer, W.~Hollik and G.~Weiglein,
Eur.\ Phys.\ J.\ C {\bf 9}, 343 (1999); 
G.~Degrassi et al., Eur.\ Phys.\ J.\ C {\bf 28}, 133 (2003).

\bibitem{vanderBij:1985ww}
J.~J.~van der Bij, Nucl.\ Phys.\ B {\bf 267} (1986) 557.

\bibitem{boos}
E. Boos et al., Nucl. Inst. Meth. A {\bf 472}, 100 (2001).

\bibitem{belusev} 
R. Belusevic, {\em Low-Energy Photon Collider}, KEK Preprint
2003-2 (2003).

\bibitem{zerwas}
M. M\"{u}hlleitner et al., DESY 00-192; hep-ph/0101083 (2001).

\bibitem{grzadkowski}
B. Grzadkowski and J. Gunion, Phys. Lett. B {\bf 294}, 361 (1992).

\bibitem{hagiwara}
K.~Hagiwara, Nucl. Instrum. Meth. A {\bf 472}, 12 (2001). 

\bibitem{Abe}
K. Abe et al., GLD Detector Outline Document (2006).

\bibitem{heinemeyer} S. Heinemeyer et al., CERN-PH-TH/2005-228; hep-ph/0511332.

\bibitem{baur}
U. Baur, T. Plehn and D.\,L. Rainwater,
Phys.\ Rev.\ D {\bf 68}, 033001 (2003).

\bibitem{belusev1}
R. Belusevic and G. Jikia, Phys. Rev. D {\bf 70}, 073017 (2004);
hep-ph/0403303.

\bibitem{Djouadi:1999gv}
A.~Djouadi et al., Eur.\ Phys.\ J.\ C {\bf 10}, 27 (1999).

\bibitem{Miller:1999ct}
D.~J.~Miller and S.~Moretti, Eur.\ Phys.\ J.\ C {\bf 13}, 459 (2000).

\bibitem{Castanier:2001sf}
C.~Castanier et al., hep-ex/0101028.

\bibitem{Belanger:2003ya}
G.~Belanger et al., Phys.\ Lett.\ B {\bf 576}, 152 (2003).

\bibitem{belusev2}
R. Belusevic, {\em A} 160--320 GeV {\em linear collider to study} $e^{+}e^{-}
\rightarrow {\rm HZ}$ {\em and} $\gamma\gamma \rightarrow {\rm H, HH}$,\\
KEK Preprint 2008-33 (2008); arXiv:0810.3187v2 (2009).

\bibitem{Jikia:1992mt}
G.~Jikia, Nucl.\ Phys.\ B {\bf 412}, 57 (1994).

\bibitem{Ginzburg:1982yr}
I.~Ginzburg et al., Nucl. Instrum. Meth. {\bf 219}, 5 (1984).

\bibitem{kawada} 
S. Kawada et al., Phys. Rev. D {\bf 85}, 113009 (2012).

\bibitem{tian}
J. Tian, K. Fujii and Y. Gao, arXiv:1008.0921v2 (2010).

\bibitem{asner}
D.~Asner et al., Eur. Phys. Journal C {\bf 28}, 27 (2003).

\bibitem{fujii}
K. Fujii, T. Matsui and Y. Sumino, Phys. Rev. D {\bf 50}, 4341 (1994).

\bibitem{ruth} R.\,D. Ruth, Proceedings of the 2001 Particle Accelerator 
Conference, Chicago.

\bibitem{adolphsen}
C.~Adolphsen, {\em Advances in Normal Conducting Accelerator Technology from
the X-band\\Linear Collider Program}, SLAC-PUB-11224, and Proceedings PAC-2005.

\bibitem{ellis}
J.~Ellis and I.~Wilson, Nature {\bf 409}, 431--435 (18 January 2001).

\bibitem{CLIC}
CLIC Conceptual Design Report, CERN 2012-XXX and KEK Report 2012-2.

\bibitem{higo}
T. Higo et al., \textit{Characteristics of Vacuum Breakdowns Under High 
Gradient Pulsed Operation}, THPS093, PASJ12, Ann. Meeting Part. Accel. Soc.
Japan, Osaka, August 2012.

\bibitem{ogitsu}
T.~Ogitsu et al., SLAC-REPRINT-1995-015 (1995). 

\bibitem{sprehn}
D.~Sprehn et al., SLAC-PUB-11162 (2004).

\bibitem{csonka}
P.~Csonka, Phys. Lett. B {\bf 24}, 625 (1967); CERN Yellow Report TH 772
(1967).

\bibitem{ginzburg}
I.~Ginzburg et al., Nucl. Instrum. Meth. {\bf 205}, 47 (1983).
\bibitem{telnovACTA}
V.~Telnov, Acta Phys. Pol. B {\bf 37}, 1049 (2006).

\bibitem{TESLA}
B. Badalek et al., Int. J. Mod. Phys. {\bf 19}, 5097 (2004).

\bibitem{saldin}
E. Saldin, E. Schneidmiller and M. Yurkov, Nucl. Instrum. Meth. A {\bf 472},
94 (2001); E. Saldin et al., SLAC-PUB-13768 (2009).

\bibitem{michelato}
P. Michelato, Proc. EPAC08, Genoa (2008); K. Fl\"{o}ttmann, private
communication.

\bibitem{ILC}
ILC Reference Design Report, G. Aarons et al. (2007).

\bibitem{skulina}
K.~Skulina et al., {\em The Future of Particle Physics}, Snowmass Conf. 2001.

\bibitem{gronberg1}
J.~Gronberg, ALC Workshop, SLAC (2004).

\bibitem{parker}
B.~Parker et al., Proc. 2007 Part. Accel. Conf., Albuquerque; 
SLAC-PUB-12832 (2007). 

\bibitem{shekhtman}
L.~Shekhtman and V.~Telnov, physics/0411253. 

\bibitem{emma}
P. Emma and T. Raubenheimer, Phys. Rev. STAB, Vol. {\bf 4}, 021001 (2001). 

\bibitem{pak}
A. Pak et al., BUDKER-INP 2003-7; hep-ex/0301037.

\bibitem{bechtel}
F. Bechtel et al., Nucl. Instrum. Meth. A {\bf 564}, 243 (2006).



\end{thebibliography}
\end{document}